\newcommand{\del}{\mbox{\boldmath{$\nabla$}}}
\newcommand{\nhat}{\mbox{\boldmath{$\hat{n}$}}}

\newcommand{\xhat}{\mbox{\boldmath{$\hat{x}$}}}

\newcommand{\phib}{\mbox{\boldmath{$\phi$}}}
\newcommand{\vb}{\mbox{{\bf v}}}

\newcommand{\Fb}{\mbox{{\bf F}}}
\newcommand{\Eb}{\mbox{{\bf E}}}
\newcommand{\ub}{\mbox{{\bf u}}}

\newcommand{\Pe}{\mbox{P\hspace{-1pt}e\hspace{2pt}}}

\newcommand{\Real}{\mbox{Re\,}}
\newcommand{\Imag}{\mbox{Im\,}}

\documentclass[12pt]{article}
\usepackage{natbib,graphicx,fullpage,doublespace,amssymb,amsfonts,psfig}
\begin{document}
\bibliographystyle{hmm}

\centerline{\large\bf 
CONFORMAL MAPPING METHODS
}
\centerline{{\large\bf 
FOR INTERFACIAL DYNAMICS}\footnote{ A perspective article to appear in
the {\it Handbook of Materials Modeling}, ed. by S. Yip et al.,
Vol. I, Ch. 4, Art. 4.10 (Springer Science and Business Media, 2005).}
}
\vspace{1em}

Microstructural evolution is typically beyond the reach of
mathematical analysis, but in two dimensions certain problems become
tractable by complex analysis.  Via the analogy between the geometry
of the plane and the algebra of complex numbers, moving free boundary
problems may be elegantly formulated in terms of conformal maps. For
over half a century, conformal mapping has been applied to continuous
interfacial dynamics, primarily in models of viscous fingering and
solidification. Current developments in materials science include
models of void electro-migration in metals, brittle fracture, and viscous
sintering.  Recently, conformal-map dynamics has also been formulated
for stochastic problems, such as diffusion-limited aggregation and
dielectric breakdown, which has reinvigorated the subject of fractal
pattern formation.

Although restricted to relatively simple models, conformal-map
dynamics offers unique advantages over other numerical methods
discussed in this chapter (such as the Level-Set Method) and in
Chapter 9 (such as the Phase Field Method). By absorbing all
geometrical complexity into a time-dependent conformal map, it is
possible to transform a moving free boundary problem to a simple,
static domain, such as a circle or square, which obviates the need for
front tracking.  Conformal mapping also allows the exact
representation of very complicated domains, which are not easily
discretized, even by the most sophisticated adaptive meshes.  Above
all, however, conformal mapping offers analytical insights for
otherwise intractable problems.

After reviewing some elementary concepts from complex analysis in \S
1, we consider the classical application of conformal mapping methods
to continous-time interfacial free boundary problems in \S 2. This
includes cases where the governing field equation is harmonic,
biharmonic, or in a more general conformally invariant class. In \S 3,
we discuss the recent use of random, iterated conformal maps to
describe analogous discrete-time phenonena of fractal growth. Although
most of our examples involve planar domains, we note in \S 4 that
interfacial dynamics can also be formulated on curved surfaces in
terms of more general conformal maps, such as stereographic
projections.  We conclude in \S 5 with some open questions
and an outlook for future research.

\section{Analytic functions and conformal maps}

We begin by reviewing some basic concepts from complex analysis found
in textbooks such as ~\cite{churchill}. For a fresh geometrical
perspective, see  ~\cite{needham}.

A general function of a complex variable depends on the real and
imaginary parts, $x$ and $y$, or, equivalently, on the linear
combinations, $z=x+iy$ and $\overline{z}=x-iy$. In contrast, an {\it
analytic function}, which is differentiable in some domain, can be
written simply as, $w=u+iv=f(z)$. The condition, $\partial f/\partial
\overline{z} = 0$, is equivalent to the Cauchy-Riemann equations,
\begin{equation}
\frac{\partial u}{\partial x} = \frac{\partial v}{\partial y}\  \mbox{
and } \
\frac{\partial u}{\partial y} = -\frac{\partial v}{\partial x},  \label{eq:CR}
\end{equation}
which follow from the existence of a unique derivative,
\begin{equation}
f^\prime = \frac{\partial f}{\partial x} =\frac{\partial u}{\partial
x} + i \frac{\partial v}{\partial x} = \frac{\partial f}{\partial
(iy)} = \frac{\partial v}{\partial y} - i \frac{\partial u}{\partial
y}, \label{eq:der}
\end{equation}
whether taken in the real or imaginary direction.

Geometrically, analytic functions correspond to a special mappings of
the complex plane. In the vicinity of any point where the derivative
is nonzero, $f^\prime(z)\neq 0$, the mapping is locally linear, $dw =
f^\prime(z) \, dz$. Therefore, an infinitessimal vector, $dz$,
centered at $z$ is transformed into another infinitessimal vector,
$dw$, centered at $w = f(z)$ by a simple complex
multiplication. Recalling Euler's formula, $(r_1 e^{i\theta_1})(r_2
e^{i\theta_2})=(r_1 r_2) e^{i(\theta_1+\theta_2)}$, this means that
the mapping causes a local stretch by $|f^\prime(z)|$ and local
rotation by $\arg f^\prime(z)$, regardless of the orientation of
$dz$. As a result, an analytic function with a nonzero derivative
describes a {\it conformal mapping} of the plane, which preserves the
angle between any pair of intersecting curves. Intuitively, a
conformal mapping smoothly warps one domain into another with no local
distortion.

Conformal mapping provides a very convenient representation of free
boundary problems. The Riemann Mapping Theorem guarantees the
existence of a unique conformal mapping between any two simply
connected domains, but the challenge is to derive its dynamics for a
given problem. The only constraint is that the conformal mapping be
{\it univalent}, or one-to-one, so that physical fields remain
single-valued in the evolving domain.

\section{Continuous interfacial dynamics}

\subsection{Harmonic fields}

Most applications of conformal mapping involve {\it harmonic}
funtions, which are solutions to Laplace's equation,
\begin{equation}
\del^2 \phi = 0 .
\end{equation}
From Eq.~(\ref{eq:CR}), it is easy to show that the real and imaginary
parts of an analytic function are harmonic, but the converse is also
true: Every harmonic is the real part of an analytic function,
$\phi=\Real \Phi$, the {\it complex potential}.  

This connection easily produces new solutions to Laplace's equation in
different geometries.  Suppose that we know the solution, $\phi(w) =
\Real \Phi(w)$, in a simply connected domain in the $w$-plane,
$\Omega_w$, which can be reached by conformal mapping, $w=f(z,t)$,
from another, possibly time-depedent domain in the $z$-plane,
$\Omega_z(t)$. A solution in $\Omega_z(t)$ is then given by 
\begin{equation}
\phi(z,t) = \Real\Phi(w) = \Real\Phi(f(z,t))   \label{eq:phitrans}
\end{equation}
because $\Phi(f(z))$ is also analytic, with a harmonic real part.  The
only caveat is that the boundary conditions be invariant under the
mapping, which holds for Dirichlet ($\phi$=constant) or Neumann
($\nhat\cdot\phi=0$) conditions.  Most other boundary conditions
invalidate Eq.~(\ref{eq:phitrans}) and thus complicate the analysis.

The complex potential is also convenient for calculating the gradient
of a harmonic function. Using Eqs.~(\ref{eq:CR}) and (\ref{eq:der}),
we have 
\begin{equation}
\nabla_z \phi = \frac{\partial \phi}{\partial x} + i \frac{\partial \phi}{\partial y}
= \overline{\Phi^\prime}, 
\end{equation}
where $\nabla_z$ is the complex gradient operator, representing the
vector gradient, $\del$, in the $z$-plane.

\noindent
{\bf{Viscous fingering and solidification:}} The classical application
of conformal-map dynamics is to {\it Laplacian growth}, where a free
boundary, $B_z(t)$, moves with a (normal) velocity,
\begin{equation}
\vb = \frac{dz}{dt} \propto \del \phi ,
\label{eq:kinematic}
\end{equation}
proportional to the gradient of a harmonic function, $\phi$, which
vanishes on the boundary (\cite{howison92}).  Conformal mapping for
Laplacian growth was introduced independently by Polubarinova-Kochina
and Galin in 1945 in the context of groundwater flow, where $\phi$ is
the pressure field and $\ub = (k/\eta) \del \phi$ is the velocity of
the fluid of viscosity, $\eta$, in a porous medium of permeability,
$k$, according to Darcy's law. Laplace's equation follows from
incompressibility, $\del\cdot\ub=0$.  The free boundary represents an
interface with a less viscous, immiscible fluid at constant pressure,
which is being forced into the more viscous fluid.

\begin{figure}[t]
\begin{center}
\mbox{
(a) \includegraphics[width=0.45\linewidth]{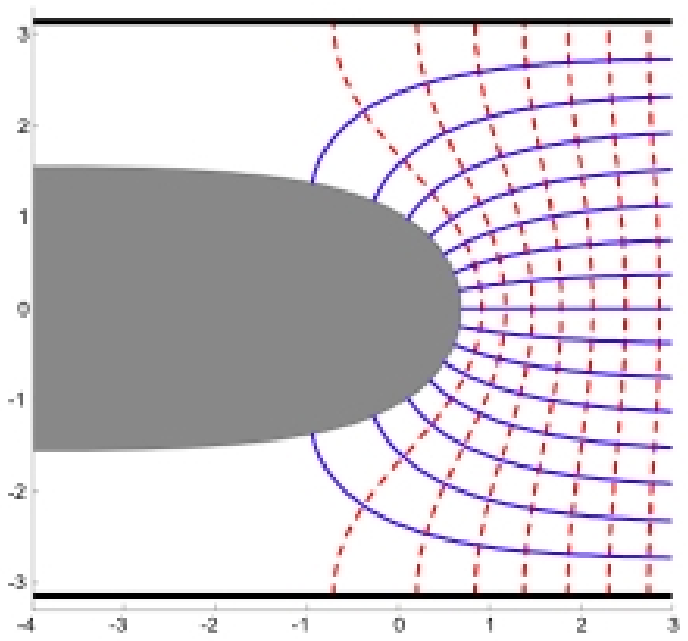} (b)
\includegraphics[width=0.45\linewidth]{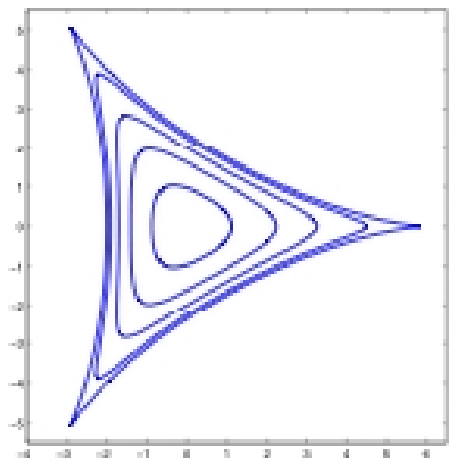} }
\caption{ \label{fig:finger}
Exact solutions for Laplacian growth, a simple model of viscous
fingering: (a) a Saffman-Taylor finger translating down an infinite
channel, showing iso-pressure curves (dashed) and streamlines (solid)
in the viscous fluid, and (b) the evolution of a perturbed circular
bubble leading to cusp singularities in finite time.  [Courtesy of
Jaehyuk Choi.] }
\end{center}
\end{figure}

In physics, Laplacian growth is viewed as a fundamental model for
pattern formation. It also describes viscous fingering in Hele-Shaw
cells, where a bubble of fluid, such as air, displaces a more viscous
fluid, such as oil, in the narrow gap between parallel flat plates.
In that case, the depth averaged velocity satisfies Darcy's law in two
dimensions.  Laplacian growth also describes dendritic solidification
in the limit of low undercooling, where $\phi$ is the temperature in
the melt (\cite{cummings99}).

To illustrate the derivation of conformal-map dynamics, let us
consider viscous fingering in a channel with inpenetrable walls, as
shown in Fig.~\ref{fig:finger}(a). The viscous fluid domain,
$\Omega_z(t)$, lies in a periodic horizontal strip, to the right of
the free boundary, $B_z(t)$, where uniform flow of
velocity, $U$, is assumed far ahead of the interface. It is convenient
to solve for the conformal map, $z = g(w,t)$, to this domain from a
half strip, $\Real w > 0$, where the pressure is simply linear, $\phi
=\Real Uw/\mu$. We also switch to dimensionless variables, where
length is scaled to a characteristic size of the initial condition,
$L$, pressure to $UL/\mu$, and time to $L/U$.

Since $\nabla_w \phi = 1$ in the half strip, the pressure
gradient at a point, $z = g(w,t)$, on the physical interface is easily
obtained from Eq.~(\ref{eq:trans}):
\begin{equation}
\nabla_z \phi = \overline{\frac{\partial f}{\partial z}} =
\left(\overline{\frac{\partial g}{\partial w}}\right)^{-1} \label{eq:fluxtrans} 
\end{equation}
where $w = f(z,t)$ is the inverse mapping (which exists as long as the
mapping remains univalent).  Now consider a Lagrangian marker, $z(t)$,
on the interface, whose pre-image, $w(t)$, lies on the imaginary axis
in the $w$-plane. Using the chain rule and Eq.~(\ref{eq:fluxtrans}),
the kinematic condition, Eq.~(\ref{eq:kinematic}), becomes,
\begin{equation}
\frac{dz}{dt} = \frac{\partial g}{\partial t} + \frac{\partial
g}{\partial w}\frac{dw}{dt} = \left(\overline{\frac{\partial
g}{\partial w}}\right)^{-1}.
\end{equation}
Multiplying by $\overline{\partial g/\partial w}\neq 0$, this
becomes
\begin{equation}
\overline{\frac{\partial g}{\partial w}}\frac{\partial g}{\partial t}
+ \left| \frac{\partial
g}{\partial w}\right|^2 \frac{dw}{dt} = 1 .
\end{equation}
Since the pre-image moves along the imaginary axis, $\Real (dw/dt) =
0$, we arrive at the {\it Polubarinova-Galin equation}  for the
conformal map:
\begin{equation}
\Real \left(  \overline{\frac{\partial g}{\partial w}}\frac{\partial
g}{\partial t} \right) = 1 , \ \ \ \mbox{ for } \Real w = 0 .   \label{eq:PG}
\end{equation}
From the solution to Eq.~(\ref{eq:PG}), the pressure is given by $\phi
= \Real f(z,t)$. Note that the interfacial dynamics is nonlinear,
even though the quasi-steady equations for $\phi$ are linear.

The best-known solutions are the Saffman-Taylor fingers,
\begin{equation}
g(w,t) = \frac{t}{\lambda} + w +2(1-\lambda) \log(1+e^{-w})
\label{eq:eqST}
\end{equation}
which translate at a constant velocity, $\lambda^{-1}$, without
changing their shape (\cite{saffman58}).  Note that (\ref{eq:eqST}) is
a solution to the fingering problem for all choices of the parameter
$\lambda$. This parameter specifies the finger width and can be chosen
arbitrarily in the solution (\ref{eq:eqST}).  In experiments however,
it is found that the viscous fingers that form are well fit by a
Saffman-Taylor finger filling precisely half of the channel, that is
with $\lambda=1/2$, as shown in Fig.~\ref{fig:finger}(a).  Why this
happens is a basic problem in {\it pattern selection}, which has been
the focus of much debate in the literature over the last 25 years.

\begin{figure}[t]
\begin{center}
\mbox{
\includegraphics[width=4in]{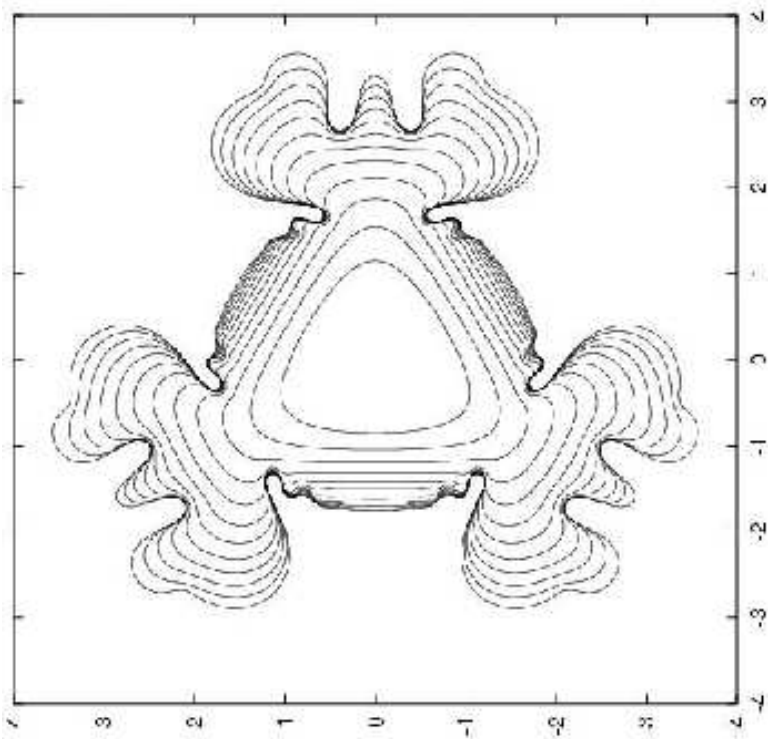} 
}
\caption{ \label{fig:siegel}
Numerical simulation of viscous fingering, starting from a three-fold
perturbation of a circular bubble. The only difference with the
Laplacian-growth dynamics in Fig.~\ref{fig:finger}(b) is the inclusion
of surface tension, which prevents the formation of cusp
singularities. [Courtesy of Michael Siegel.]
}
\end{center}
\end{figure}
\nopagebreak 

To understand this problem, note that the viscous finger solutions
(\ref{eq:eqST}) do not include any of the effects of surface tension
on the interface between the two fluids.  The intriguing pattern
selection of the $\lambda=1/2$ finger has been attributed to a
singular perturbation effect of small surface tension.  Surface
tension, $\gamma$, is a significant complication because it is
described by a non-conformally-invariant boundary condition,
\begin{equation}
\phi = \gamma\kappa , \ \ \mbox{ for } z\in B_z(t)
\end{equation}
where $\kappa$ is the local interfacial curvature, entering via the
Young-Laplace pressure.  Small surface tension can be treated
analytically as a singular perturbation to gain insights into pattern
selection (\cite{kruskal91,tanveer93}).  Since surface tension
effects are only significant at points of high curvature $\kappa$ in
the interface, and given that the finger in Fig. \ref{fig:finger}(a)
is very smooth with no such points of high curvature, it is surprising
that surface tension acts to select the finger width. Indeed, the
viscous fingering problem has been shown to be full of surprises
(\cite{surprise}).

In a radial geometry, the univalent mapping is from the exterior of
the unit circle, $|w|=1$, to the exterior of a finite bubble
penetrating an infinite viscous liquid. \cite{ben84} introduced a {\it
pole dynamics} formulation, where the map is expressed in terms of its
zeros and poles, which must lie inside the unit circle to preserve
univalency. They showed that Laplacian growth in this geometry is
ill-posed, in the sense that cusp-like singularites occur in finite
time (as a zero hits the unit circle) for a broad class of initial
conditions, as illustrated in Fig.~\ref{fig:finger}(b). (See
\cite{howison92} for a simple, general proof due to Hohlov.) This
initiated a large body of work on how Laplacian growth is
``regularized'' by surface tension or other effects in real systems.

Despite the analytical complications introduced by surface tension,
several exact steady solutions with non-zero surface tension are known
(\cite{kadanoff90,crowdy00}).  Surface tension can also be
incorporated into numerical simulations based on the same
conformal-mapping formalism (\cite{maclean81}), which show how cusps
are avoided by the formation of new fingers (\cite{dai91}). For
example, consider a three-fold perturbation of a circular bubble,
whose exact dynamics without surface tension is shown in
Fig.~\ref{fig:finger}(b). With surface tension included, the evolution
is very similar until the cusps begin to form, at which point the tips
bulge outward and split into new fingers, as shown in
Fig.~\ref{fig:siegel}. This process repeats itself to produce a
complicated fractal pattern (\cite{bunde}), which curiously ressembles
the diffusion-limited particle aggregates discussed below in \S 3.

\noindent
{\bf{Density-driven instabilities in fluids:}} An important class of
problems in fluid mechanics involves the nonlinear dynamics of an
interface between two immiscible fluids of different densities. In the
presence of gravity, there are some familiar cases.  Deep-water waves
involve finite disturbances (such as steady ``Stokes waves''') in the
interface between lighter fluid (air) over a heavier fluid
(water). With an inverted density gradient, the Rayleigh-Taylor
instability develops when a heavier fluid lies above a lighter fluid,
leading to large plumes of the former sinking into the
latter. \cite{tan93} has used conformal mapping 
to analye the Rayleigh-Taylor instability and has provided evidence to
associate the formation of plumes with the approach of various
conformal mapping singularities to the unit circle.

\begin{figure}[ht]
\begin{center}
\mbox{
\includegraphics[width=4in]{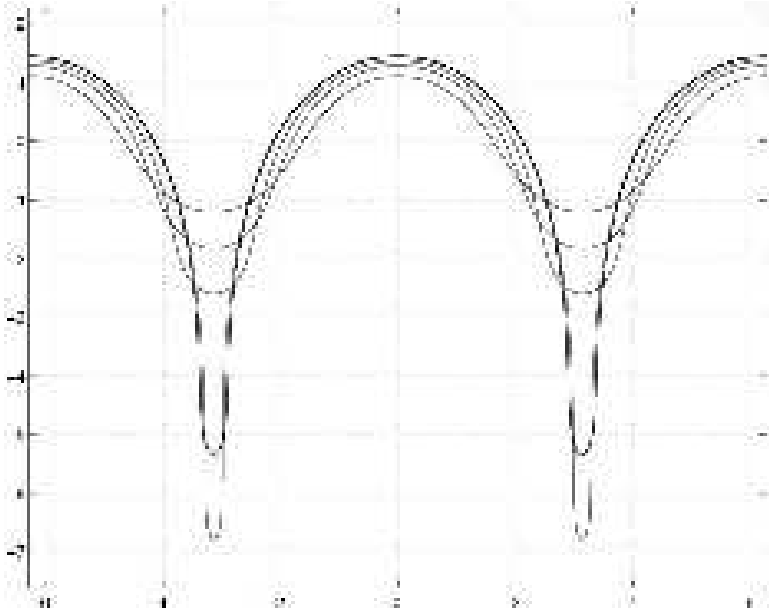} 
}
\caption{ \label{fig:toshio}
Conformal-map dynamics for the strongly nonlinear regime of the
Richtmyer-Meshkov instability (\cite{yoshi99}). [Courtesy
of Toshio Yoshikawa and Alexander Balk.]  }
\end{center}
\end{figure}

\nopagebreak

A related problem is the Richtmyer-Meshkov instability, which occurs
when a shock wave passes through an interface between fluids of
different densities. Behind the shock, narrow fingers of the heavier
fluid penetrate into the lighter fluid. The shock wave usually passes
so quickly that compressibility only affects the onset of the
instability, while the nonlinear evolution occurs much faster than the
development of viscous effects. Therefore, it is reasonable to assume
a potential flow in each fluid region, with randomly perturbed initial
velocities. Although real plumes roll up in three dimensions and
bubbles can form, conformal mapping in two dimensions still
provides some insights, with direct relevance for shock tubes of high
aspect ratio.

A simple conformal-mapping analysis is possible
for the case of a large density contrast, where the lighter fluid is
assumed to be at uniform pressure. The
Richtmyer-Meshkov instability (zero-gravity limit) is then similar to
the Saffman-Taylor instability, except that the total volume of each
fluid is fixed.  A periodic interface in the $y$ direction, analogous
to the channel geometry in Fig.~\ref{fig:finger}, can be described by
the univalent mapping, $z=g(w,t)$, from the interior of the unit
circle in the mathematical $w$ plane to the interior of the
heavy-fluid finger in the physical $z$ plane.  

\cite{zakharov68} introduced a Hamiltonian formulation of
the interfacial dynamics in terms of this conformal map, taking into
account kinetic and potential energy, but not surface tension.  One
way to derive equations of motion is to expand the map in a {\it
Taylor series},
\begin{equation}
g(w,t) = \log w + \sum_{n=0}^\infty a_n(t) w^n, \ \ \ |w|<1 .
\label{eq:taylor}
\end{equation}
(The $\log w$ term first maps the disk to a periodic half strip.) On
the unit circle, $w=e^{i\theta}$, the pre-image of the interface,
this is simply a complex Fourier series. The Taylor
coefficients, $a_n(t)$, act generalized coordinates describing
$n$-fold shape perturbations within each period, and their time
derivatives, $\dot{a}_n(t)$, act as velocities or momenta.
Unfortunately, truncating the Taylor series results in a poor
description of strongly nonlinear dynamics because the conformal map
begins to vary wildly near the unit circle. 

An alternate approach used by \cite{yoshi99} is to expand in terms
ressembling Saffman-Taylor fingers,
\begin{equation}
g(w,t) = \log w + b(t) - \sum_{n=1}^N b_n(t) \log(1-\lambda_n(t) w) \label{eq:fingers}
\end{equation}
which can be viewed as a resummation of the Taylor series in
Eq.~(\ref{eq:taylor}). As shown in Fig.~\ref{fig:toshio}, exact
solutions exist with only a finite number terms in the finger
expansion, as long as the new generalized coordinates, $\lambda_n(t)$,
stay inside the unit disk, $|\lambda_n|<1$.
This example illustrates the importance of the choice of {\it shape
functions} in the expansion of the conformal map, e.g.  $w^n$ versus
$\log(1-\lambda_n w)$.

\noindent
{\bf{Void electro-migration in metals:}} Micron-scale interconnects in
modern integrated circuits, typically made of aluminum, sustain
enormous currents and high temperatures. The intense electron wind
drives solid-state mass diffusion, especially along dislocations and
grain boundaries, where voids also nucleate and grow. In the narrowest
and most fragile interconnects, grain boundaries are often well
separated enough that isolated voids migrate in a fairly homogeneous
environment due to surface diffusion, driven by the electron wind.
Voids tend to deform into slits, which propagate across the
interconnect, causing it to sever.  A theory of void electro-migration
is thus important for predicting reliability.
\begin{figure}
\begin{center}
\mbox{
(a) \includegraphics[width=0.45\linewidth]{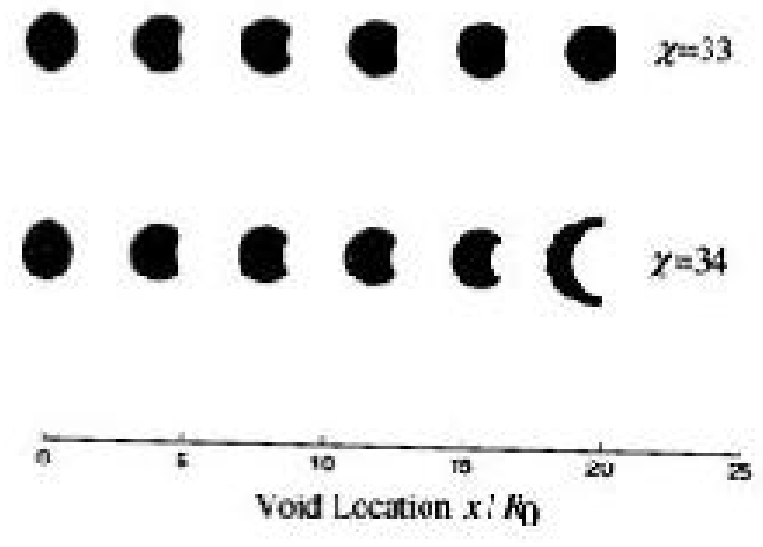}
(b) \includegraphics[width=0.45\linewidth]{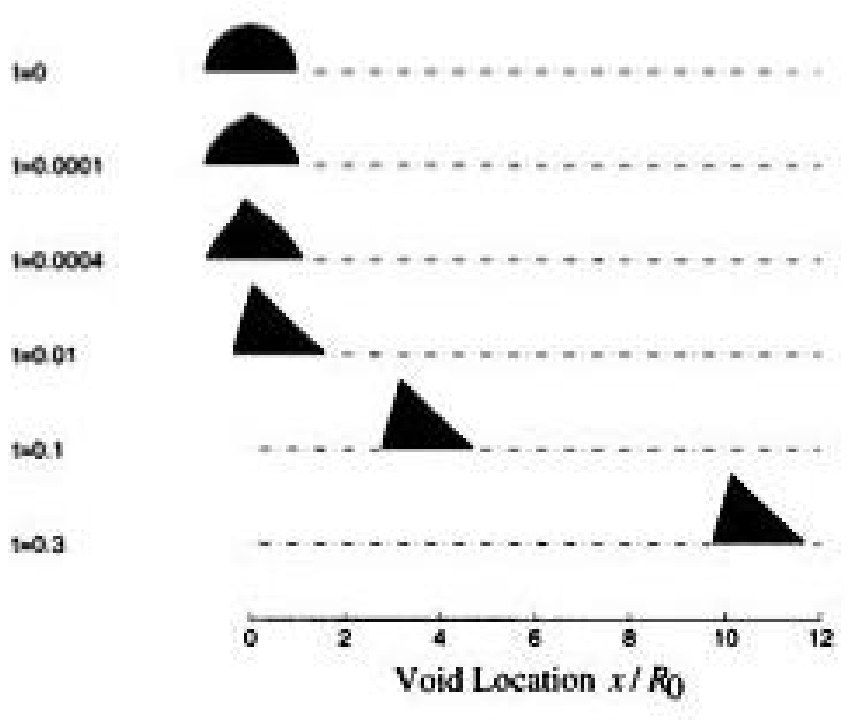}
}
\caption{ \label{fig:void}
Numerical conformal-mapping simulations of the electromigration of
voids in aluminum interconnects (\cite{wang96}). (a) A small shape
perturbation of a cylindrical void decaying (above) or deforming into
a curved slit (below), depending on a dimensionless group, $\chi$,
comparing the electron wind to surface-tension gradients. (b) A void
evolving with anisotropic surface diffusivity ($\chi=100, g_d=100,
m=3$).  [Courtesy of Zhigang Suo.] }
\end{center}
\end{figure}

In the simplest two-dimensional model (\cite{wang96}), a cylindrical
void is modeled as a deformable, insulating inclusion in a conducting
matrix. Outside the void, the electrostatic potential, $\phi$,
satisfies Laplace's equation, which invites the use of conformal
mapping. The electric field, $\Eb = -\del\phi$, is taken to be uniform
far away and wraps around the void surface, due to a Neuman boundary
condition, $\nhat\cdot\Eb=0$.

The difference with Laplacian growth lies in the kinematic condition,
which is considerably more complicated. In place of
Eq.~(\ref{eq:kinematic}), the normal velocity of the void surface is
given by the surface divergence of the surface current, $j$, which
takes the dimensionless form,
\begin{equation}
\nhat\cdot\vb = \frac{\partial j}{\partial s} =
\chi \frac{\partial^2\phi}{\partial s^2}
+  \frac{\partial^2\kappa}{\partial s^2}
\end{equation}
where $s$ is the local arc-length coordinate and $\chi$ is a
dimensionless parameter comparing surface currents due to the electron
wind force (first term) and due to gradients in surface tension
(second term). This moving free boundary problem somewhat resembles
the viscous fingering problem with surface tension, and it admits
analogous finger solutions, albeit of width $2/3$, not $1/2$
(\cite{benamar99}).

To describe the evolution of a singly connected void, we consider the
conformal map, $z=g(w,t)$, from the exterior of the unit circle to the
exterior of the void. As long as the map remains univalent, it has a
{\it Laurent series} of the form,
\begin{equation}
g(w,t) = A_1(t) w + A_0(t) + \sum_{n=1}^\infty A_{-n}(t) w^{-n},  \ \ \ \mbox{
for} |w|>1 ,  \label{eq:laurent}
\end{equation}
where the Laurent coefficients, $A_n(t)$, are now the generalized
coordinates.  As in the case of viscous fingering (\cite{howison92}),
a hierarchy of nonlinear ordinary differential equations (ODEs) for these
coordinates can be derived. For void electromigration, \cite{wang96}
start from a variational principle accounting for surface tension and
surface diffusion, using a Galerkin procedure. They truncate the
expansion after $17$ coefficients, so their numerical method breaks
down if the void deforms significantly, e.g. into curved
slit. Nevertheless, as shown in Fig,.~\ref{fig:void}(a), the numerical
method is able to capture essential features of the early stages of
strongly nonlinear dynamics.

In the same regime, it is also possible to incorporate anisotropic
surface tension or surface mobility. The latter involves multiplying
the surface current by a factor $(1+g_d \cos m\alpha)$, where $\alpha$
is the surface orientation in the physical $z$ plane, given at $z =
g(e^{i\theta},t)$, by 
\begin{equation}
\alpha = \theta + \arg \frac{\partial g}{\partial w}(e^{i\theta},t) .
\label{eq:aniso}
\end{equation}
Some results are shown in Fig.~\ref{fig:void}(b), where the void
develops dynamical facets.

\noindent {\bf Quadrature domains:~}
We end this section by commenting on some of the mathematics
underlying the existence of exact solutions to continuous-time
Laplacian-growth problems.  Significantly, much of this mathematics
carries over to problems in which the governing field equation is not
necessarily harmonic, as will be seen in the following section.

The steadily-translating finger solution (\ref{eq:eqST}) of Saffman
and Taylor turns out to be but one of an increasingly large number of
known exact solutions to the standard Hele-Shaw
problem. \cite{saffman59} himself identified a class of unsteady
finger-like solutions. This solution was later generalized by
\cite{howison86} to solutions involving multiple fingers exhibiting
such phenomena as {\it tip-splitting} where a single finger splits
into two (or more) fingers. It is even possible to find exact
solutions to the more realistic case where there is a second interface
further down the channel (\cite{crowdy04}) which must always be the
case in any experiment.

Besides finger-like solutions which are characterized by time-evolving
conformal mappings having logarithmic branch-point singularities,
other exact solutions, where the conformal mappings are rational
functions with time-evolving poles and zeros, were first identified by
Polubarinova-Kochina and Galin in 1945. \cite{Rich81} later rediscovered
the latter solutions while simultaneously presenting important new
theoretical connections between the Hele-Shaw problem and a class of
planar domains known as {\it quadrature domains}.  The simplest
example of a quadrature domain is a circular disc $D$ of radius $r$
centred at the origin which satisfies the identity
\begin{equation}
\int \int_{D} h(z) dx dy = \pi r^2 h(0)
\label{eq:eqMVT}
\end{equation}
where $h(z)$ is any function analytic in the disc (and integrable over
it).  Equation (\ref{eq:eqMVT}), which is known as a {\it quadrature
identity} since it holds for any analytic function $h(z)$, is simply a
statement of the well-known {\it mean-value theorem} of complex
analysis (\cite{carrier}).  A more general domain $D$, satisfying a
generalized quadrature identity of the form
\begin{equation}
\int \int_{D} h(z) dx dy = \sum_{k=1}^{N} \sum_{j=0}^{n_k-1} c_{jk} h^{(j)}(z_k)
\label{eq:eqQD}
\end{equation} 
is known as a quadrature domain. Here, $\lbrace z_k \in \mathbb{C}
\rbrace$ is a set of points inside $D$ and $h^{(j)}(z)$ denotes the
$j$-th derivative of $h(z)$.  If one makes the choice $h(z)=z^n$ in
(\ref{eq:eqQD}) the resulting integral quantities have become known as
the {\it Richardson moments} of the domain.  Richardson showed that
the dynamics of the Hele-Shaw problem is such as to {\it preserve}
quadrature domains. That is, if the initial fluid domain in a
Hele-Shaw cell is a quadrature domain at time $t=0$, it remains a
quadrature domain at later times (so long as the solution does not
break down).  This result is highly significant and provides a link
with many other areas of mathematics including potential theory, the
notion of balayage, algebraic curves, Schwarz functions and Cauchy
transforms. \cite{Rich92} discusses many of these connections while
\cite{var92} provide a more general overview of
the various mathematical implications of Richardson's result.
\cite{shap92} gives more general background on quadrature domain theory.

It is a well-known result in the theory of quadrature domains
(\cite{shap92}) that simply-connected quadrature domains can be
parametrized by rational function conformal mappings from a unit
circle. Given Richardson's result on the preservation of quadrature
domains, this explains why Polubarinova-Kochina and Galin were able
to find time-evolving rational function conformal mapping solutions to
the Hele-Shaw problem.  It also underlies the pole dynamics results of
\cite{ben84}.  But Richardson's result is not restricted to
simply-connected domains; {\it multiply-connected} quadrature domains
are also preserved by the dynamics.  Physically this corresponds to
time-evolving fluid domains containing multiple bubbles of air.
Indeed, motivated by such matters, recent research has focused on the
analytical construction of multiply-connected quadrature domains using
conformal mapping ideas (\cite{Rich01}, \cite{CroMar}). In the
higher-connected case, the conformal mappings are no longer simply
rational functions but are given by conformal maps that are {\it
automorphic functions} (or, meromorphic functions on compact Riemann
surfaces).  The important point here is that understanding the
physical problem from the more general perspective of quadrature
domain theory has led the way to the unveiling of more sophisticated
classes of exact conformal mapping solutions.

\subsection{Bi-Harmonic fields}

Although not as well known as conformal mapping of harmonic functions,
there is also a substantial literature on complex-variable methods to
solve the {\it bi-harmonic equation},
\begin{equation}
\del^2 \del^2 \psi = 0 , \label{eq:bi}
\end{equation}
which arises in two-dimensional elasticity (\cite{musk}) and fluid
mechanics (\cite{batchelor}).  Unlike harmonic functions, which can be expressed in terms
of a single analytic function (the complex potential), bi-harmonic
functions can be expressed in terms of two analytic functions, $f(z)$
and $g(z)$, in {\it Goursat form} (\cite{carrier}):
\begin{equation}
\psi(z, \bar z) = \Imag [\bar z f(z) + g(z)] \label{eq:goursat}
\label{eq:eqGoursat}
\end{equation}
Note that $\psi$ is no longer just the imaginary part of an analytic
function $g(z)$ but also contains the imaginary part of the
non-analytic component $\bar z f(z)$.  A difficulty with bi-harmonic
problems is that the typical boundary conditions (see below) are not
conformally invariant, so conformal mapping does not usually generate
new solutions by simply a change of variables, as in
Eq.~(\ref{eq:phitrans}). Nevertheless, the Goursat form of the
solution, Eq.~(\ref{eq:goursat}), is a major simplification, which
enables analytical progress.

\noindent {\bf{Viscous sintering:}}
Sintering describes a process by which a granular compact of particles
(e.g. metal or glass) is raised to a sufficiently large temperature
that the individual particles become mobile and release surface energy
in such a way as to produce inter-particulate bonds. At the start of a
sinter process, any two particles which are initially touching develop
a thin ``neck'' which, as time evolves, grows in size to form a more
developed bond. In compacts in which the packing is such that
particles have more than one touching neighbour, as the necks grow in
size, the sinter body densifies and any enclosed pores between
particles tend to close up.  The macroscopic material properties of
the compact at the end of the sinter process depend heavily on the
degree of densification.  In industrial application, it is crucial to
be able to obtain accurate and reliable estimates of the time taken
for pores to close (or reduce to a sufficiently small size) within any
given initial sinter body in order that industrial sinter times are
optimized without compromising the macroscopic properties of the final
densified sinter body.

\begin{figure}
\begin{center}
\mbox{
\includegraphics[width=0.9\linewidth]{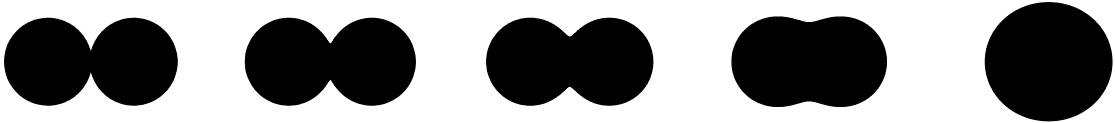}
}
\caption{Evolution of the solution of \cite{Hopper} for the
coalescence of two equal blobs of fluid under the effects of surface
tension. \label{fig:Hopper}}
\end{center}
\end{figure}

The fluid is modelled as a region $D(t)$ of
very viscous, incompressible fluid, in which the velocity field,
\begin{equation}
{\bf u} = (u,v) = (\psi_y, -\psi_x).
\end{equation}
is given by the curl of an out-of-plane vector, whose magnitude is a
stream function, $\psi(x,y,t)$, which satisfies the bi-harmonic
equation (\cite{batchelor}).  On the boundary $\partial D(t)$, the
tangential stress must vanish and the normal stress must be balanced
by the uniform surface tension effect, i.e.,
\begin{equation}
-p n_i + 2 \mu e_{ij} = T \kappa n_i,  \label{eq:stress}
\end{equation}
where $p$ is the fluid pressure, $\mu$ is the viscosity, $T$ is the
surface tension parameter, $\kappa$ is the boundary curvature, $n_i$
denotes components of the outward normal ${\bf n}$ to $\partial D(t)$
and $e_{ij}$ is the usual fluid rate-of-strain tensor.  The boundary
is time-evolved in a quasi-steady fashion with a normal velocity,
$V_n$, determined by the same kinematic condition, $V_n = {\bf u}\cdot{\bf
n}$, as in viscous fingering.

In terms of the Goursat functions in (\ref{eq:goursat}) -- which are
now generally time-evolving -- the stress condition (\ref{eq:stress})
takes the form
\begin{equation}
f(z,t) + z 
\overline{f'}(\bar z,t) + 
\overline{g'}(\bar z,t) = -{i \over 2} z_s
\label{eq:eq2}
\end{equation} 
where again $s$ denotes arclength.
Once $f(z,t)$ has been determined from (\ref{eq:eq2}),
the kinematic condition
\begin{equation}
{\rm Im}[z_t \bar z_s] = 
{\rm Im}[- 2 f(z,t) \bar z_s] -{1 \over 2}
\end{equation}
is used to time-advance the interface.

A significant contribution was made by \cite{Hopper} who showed, using
complex variable methods based on the decomposition
(\ref{eq:eqGoursat}), that the problem for the surface-tension driven
coalescence of two equal circular blobs of viscous fluid can be
reduced to the evolution of a rational function conformal map, from a
unit $w$-circle, of the form
\begin{equation}
g(w,t) = {R(t) w \over w^2 - a^2(t)}.
\label{eq:eqHop}
\end{equation}
The two time-evolving parameters $R(t)$ and $a(t)$ satisfy two coupled
nonlinear ODEs.  Figure \ref{fig:Hopper} shows a sequence of shapes of
the two coalescing blobs computed using Hopper's solution.  At large
times, the configuration equilibrates to a single circular blob.

\begin{figure}
\begin{center}
\mbox{
\includegraphics[width=0.5\linewidth]{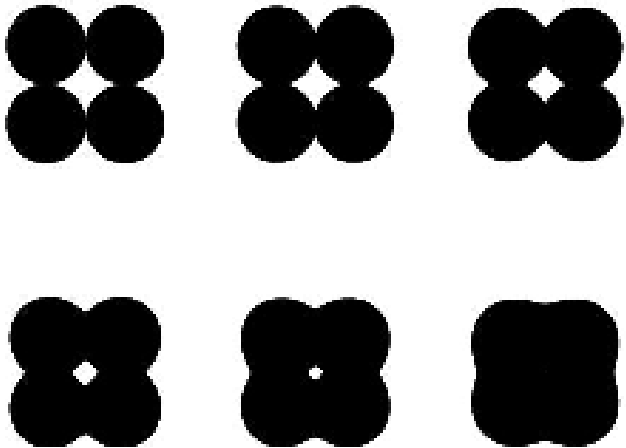}
}
\caption{ \label{fig:sintering} The coalescence of fluid blobs
and collapse of cylindrical pores in a model of viscous sintering. This
sequence of images shows an analytical solution by \cite{crowdy03}
using complex-variable methods. }
\end{center}
\end{figure}

While Hopper's coalescence problem provides insight into the growth of
the interparticle neck region, there are no pores in this
configuration and it is natural to ask whether more general exact
solutions exist. \cite{crowdy99} reappraised the viscous sintering
problem and showed, in direct analogy with Richardson's result on
Hele-Shaw flows, that the dynamics of the sintering problem is also
such as to preserve quadrature domains.  As in the Hele-Shaw problem,
this perspective paved the way for the identification of new exact
solutions, generalizing (\ref{eq:eqHop}), for the evolution of
doubly-connected fluid regions.  Figure \ref{fig:sintering} shows the
shrinkage of a pore enclosed by a typical ``unit'' in a
doubly-connected square packing of touching near-circular blobs of
viscous fluid. This calculation employs a conformal mapping to the
doubly-connected fluid region (which is no longer a rational function
but a more general automorphic function) derived by
\cite{crowdy03} and, in the same spirit as Hopper's solution (\ref{eq:eqHop}),
requires only the integration of three coupled nonlinear ODEs. The
fluid regions shown in Figure \ref{fig:sintering} are all
doubly-connected quadrature domains.
\cite{Rich00} has also considered similar Stokes flow problems using
a different conformal mapping approach.

\noindent
{\bf{Pores in elastic solids:}} Solid elasticity in two dimensions is
also governed by a bi-harmonic function, the Airy stress function
(\cite{musk}).  Therefore, the stress tensor, $\sigma_{ij}$, and the
displacement field, $u_i$, may be expressed in terms of two analytic
functions, $f(z)$ and $g(z)$:
\begin{eqnarray}
\frac{\sigma_{22} + \sigma_{11}}{2} &=& f^\prime(z) +
\overline{f^\prime}(\overline{z}) \\
\frac{\sigma_{22}-\sigma_{11}}{2} + i \sigma_{12} &=& \overline{z}
f^{\prime\prime}(z) + g^\prime(z) \\
\frac{Y}{1+\nu} (u_1 + i u_2) &=& \kappa f(z) - z
\overline{f^\prime}(\overline{z}) - \overline{g}(\overline{z})
\end{eqnarray}
where $Y$ is Young's modulus, $\nu$ is Poisson's ratio, and $\kappa =
(3-\nu)/(1+\nu)$ for plane stress and $\kappa = 3-4\nu$ for plane
strain.  As with bubbles in viscous flow, the use of Goursat
functions allows conformal mapping to be applied to bi-harmonic free
boundary problems in elastic solids, without solving
explicitly for bulk stresses and strains.

For example, \cite{wang97} have simulated the dynamics of a
singly-connected pore by surface diffusion in an infinite stressed
elastic solid.  As in the case of void electromigration described
above, they solve nonlinear ODEs for the Laurent coefficients of the
conformal map from the exterior of the unit disk,
Eq.~(\ref{eq:laurent}).  Under uniaxial tension, there is a
competition between surface tension, which prefers a circular shape,
and the applied stress, which drives elongation and eventually
fracture in the transverse direction. The numerical method, based on
the truncated Laurent expansion, is able to capture the transition
from stable elliptical shapes at small applied stress to the unstable
growth of transverse protrusions at large applied stress, although
naturally it breaks down when cusps ressembling crack tips begin to
form.

\subsection{ Non-harmonic conformally invariant fields }

The vast majority of applications of conformal mapping fall into one
of the two classes above, involving harmonic or bi-harmonic functions,
where the connections with analytic functions,
Eqs.~(\ref{eq:phitrans}) and (\ref{eq:goursat}), are cleverly
exploited.  It turns out, however, that conformal mapping can be
applied just as easily to a broad class of problems involving
non-harmonic fields, recently discovered by \cite{bazant04}. Of
course, in planar geometry, the conformal map itself is described by
an analytic function, but the fields need not be, as long as they
transform in a simple way under conformal mapping.

The most convenient fields satisfy {\it conformally invariant} partial
differential equations (PDEs), whose forms are unaltered by a
conformal change of variables. It is straightforward to transform PDEs
under a conformal mapping of the plane, $w=f(z)$, by expressing them
in terms of complex gradient operator introduced above,
\begin{equation}
\nabla_z  = \frac{\partial}{\partial x} + i \frac{\partial}{\partial y} = 2
\frac{\partial}{\partial \overline{z}} , \label{eq:trans}
\end{equation}
which we have related to the $\overline{z}$ partial derivative using
the Cauchy-Riemann equations, Eq.~(\ref{eq:CR}).  In this form, it is
clear that $\nabla_z f = 0$ if and only if $f(z)$ is analytic, in
which case $\overline{\nabla}_z f = 2 f^\prime$. Using the chain rule, also obtain the
transformation rule for the gradient, 
\begin{equation}
\nabla_z = \overline{f^\prime} \, \nabla_w  \label{eq:gradtrans}
\end{equation}

To apply this formalism, we write Laplace's equation in the form,
\begin{equation}
\del_z^2\phi = \Real \nabla_z \overline{\nabla}_z \phi = \nabla_z \overline{\nabla}_z
\phi = 0 ,
\end{equation}
which assumes that mixed partial derivatives can be taken in either
order. (Note that $\mbox{\bf a}\cdot\mbox{\bf b} = \Real a
\overline{b}$.) The conformal invariance of Laplace's equation,
$\nabla_w\overline{\nabla}_w \phi = 0$, then follows from a simple
calculation,
\begin{equation}
\nabla_z \overline{\nabla}_z  
= (\nabla_z f^\prime)\overline{\nabla}_w + |f^\prime|^2 \nabla_w
\overline{\nabla}_w =  |f^\prime|^2 \, \nabla_w  \overline{\nabla}_w
\label{eq:lapl}
\end{equation}
where $\nabla_z f^\prime =0$ because $f^\prime$ is also analytic. As a
result of conformal invariance, any harmonic function in the $w$
plane, $\phi(w)$, remains harmonic in the $z$ plane, $\phi(f(z))$,
after the simple substitution, $w=f(z)$. We came to the same
conclusion above in Eq.~(\ref{eq:phitrans}), using the connection
between harmonic and analytic functions, but the argument here is more
general and also applies to other PDEs.

The bi-harmonic equation is not conformally invariant, but some other
equations -- and systems of equations -- are.  The key observation is
that any ``product of two gradients'' transforms in the same way under
conformal mapping, not only the Laplacian, $\del\cdot\del
\phi$, but also the term, $\del\phi_1\cdot\del \phi_2 = \Real (\nabla\phi_1)
\overline{\nabla} \phi_2$, which involves {\it two} real functions, $\phi_1$ and $\phi_2$:
\begin{equation}
\Real \, (\nabla_z \phi_1)\, \overline{\nabla}_z \phi_2 = |f^\prime|^2 \, \Real \,
(\nabla_w \phi_1)\, \overline{\nabla}_w \phi_2. \label{eq:conv}
\end{equation}
(Todd Squires has since noted that the operator, $\del \phi_1
\times \del \phi_2= \Imag (\nabla\phi_1)\overline{\nabla \phi_2}$, also transforms
in the same way.) These observations imply the conformal invariance of
a broad class of systems of nonlinear PDEs:
\begin{equation}
\sum_{i=1}^N \left( \, a_i \,\del^2 \phi_i + \sum_{j=i}^N 
\,a_{ij}\,\del \phi_i \cdot \del \phi_j\, + \sum_{j=i+1}^N \,b_{ij}\,
\del\phi_i \times \del\phi_j \, \right) = 0
\label{eq:geneq}
\end{equation}
where the coefficients $a_i(\phib)$, $a_{ij}(\phib)$, and
$b_{ij}(\phib)$ may be nonlinear functions of the unknowns, $\phib =
(\phi_1, \phi_2, \ldots, \phi_N)$, but not of the independent
variables or any derivatives of the unknowns.

The general solutions to these equations are not harmonic and thus
depend on both $z$ and $\overline{z}$. Nevertheless, conformal mapping
works in precisely the same way: A solution, $\phib(w,\overline{w})$,
can be mapped to another solution, $\phib(f(z),\overline{f(z)})$, by a
simple substitution, $w=f(z)$. This allows the conformal mapping
techniques above (and below) to be extended to new kinds of moving
free boundary problems.

\noindent {\bf Transport-limited growth phenomena:}
For physical applications, the conformally invariant class,
Eq.~(\ref{eq:geneq}), includes the important set of steady
conservation laws for gradient-driven flux densities,
\begin{equation}
\frac{\partial c_i}{\partial t} = \del\cdot\Fb_i = 0, \ \ \ 
\Fb_i = c_i\, \ub_i - D_i(c_i)\, \del c_i, \ \ \  \ub_i \propto
\del \phi  \label{eq:Fi}
\end{equation}
where $\{c_i\}$ are scalar fields, such as chemical concentrations or
temperature, $\{D_i(c_i)\}$ are nonlinear diffusivities, $\{\ub_i\}$
are irrotational vector fields causing advection, and $\phi$ is a
potential (\cite{bazant04}). Physical examples include
advection-diffusion, where $\phi$ is the harmonic velocity potential,
and electrochemical transport, where $\phi$ is the non-harmonic
electrostatic potential, determined implicitly by electro-neutrality.

By modifying the classical methods described above for
Laplacian growth, conformal-map dynamics can thus be formulated for
more general, transport-limited growth phenomena (\cite{ADLA}).  The
natural generalization of the kinematic condition,
Eq.~(\ref{eq:kinematic}), is that the free boundary moves in
proportion to one of the gradient-driven fluxes with velocity, $\vb
\propto \Fb_1$. For the growth of a finite filament, driven by
prescribed fluxes and/or concentrations at infinity, one obtains a
generalization of the Polubarinova-Galin equation for the conformal
map, $z = g(w,t)$, from the exterior of the unit disk to the exterior
of growing object,
\begin{equation}
\Real (\overline{w\, g^\prime} \, g_t) = \sigma(w,t), \ \ \mbox{ on }
|w|=1 \label{eq:genPB}
\end{equation}
where $\sigma(w,t)$ is the non-constant, time-dependent normal flux,
$\nhat\cdot\Fb_1$, on the unit circle in the mathematical plane. 

\begin{figure}
\begin{center}
\mbox{
\includegraphics[width=0.75\linewidth]{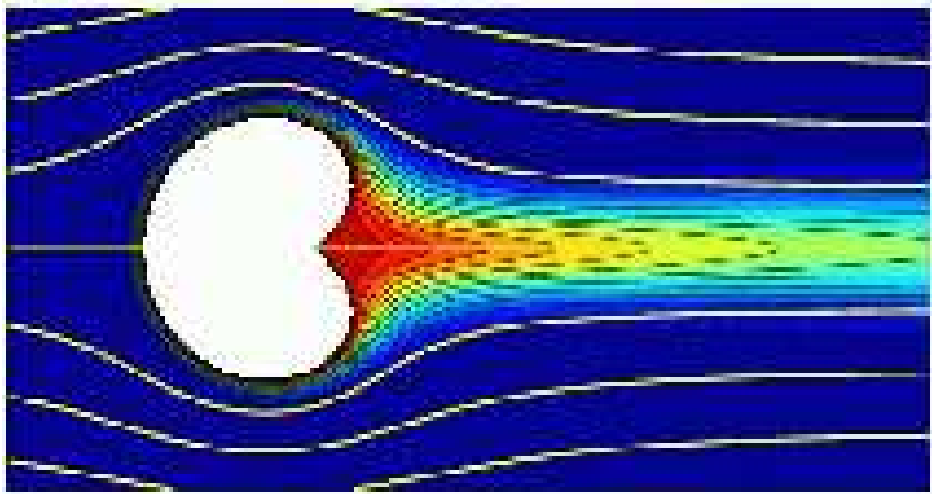}
}
\caption{  
The exact self-similar solution, Eq.~(\ref{eq:adfp}), for continuous
advection-diffusion-limited growth in a uniform background potential
flow (yellow streamlines) at the dynamical fixed point
($\Pe=\infty$). The concentration field (color contour plot) is shown
for $\Pe=100$. [Courtesy of Jaehyuk Choi.]
\label{fig:ADLAcontinuum}
}
\end{center}
\end{figure}

\noindent
{\bf{Solidification in a fluid flow:}} A special case of the
conformally invariant equations (\ref{eq:geneq}) has been known for
almost a century: steady advection-diffusion of a scalar field, $c$,
in a potential flow, $\ub$. The dimensionless PDEs are
\begin{equation}
\Pe\, \ub\cdot\del c =  \del^2 c , \ \ \ 
\label{eq:ad}
\ub = \del\phi, \ \ \ \del^2 \phi=0
\end{equation}
where we have introduced the P\'eclet number, $\Pe = UL/D$, in terms
of a characteristic length, $L$, velocity, $U$, and diffusivity, $D$.
In 1905, Boussinesq showed that Equation~(\ref{eq:ad}) takes a simpler
form in {\it streamline coordinates}, $(\phi,\psi)$, where $\Phi =
\phi + i\psi$ is the complex velocity potential:
\begin{equation}
\Pe\, \frac{\partial c}{\partial \phi} = \left( \frac{\partial^2 c}{\partial
\phi^2} + \frac{\partial^2 c}{\partial \psi^2} \right)
\end{equation}
because advection (the left hand side) is directed only along
streamlines, while diffusion (the right hand side) also occurs in the
transverse direction, along isopotential lines. From the general
perspective above, we recognize this as the conformal mapping of an
invariant system of PDEs of the form (\ref{eq:Fi}) to the complex $\Phi$
plane, where the flow is uniform and any obstacles in the flow are
mapped to horizontal slits.

Streamline coordinates form the basis for Maksimov's method for
interfacial growth by steady advection-diffusion in a background
potential flow, which has been applied to freezing in groundwater flow
and vapor deposition on textile fibers
(\cite{kornev94,cummings99}). The growing interface is a streamline
held at a fixed concentration (or temperature) relative to the flowing
bulk fluid at infinity. This is arguably the simplest growth model
with two competing transport processes, and yet open questions remain
about the  nonlinear dynamics, even without surface tension.

The normal flux distribution to a finite absorber in a uniform
background flow, $\sigma(w,t)$ in Eq.~(\ref{eq:genPB}) is well known,
but rather complicated (\cite{advdif}), so it is replaced by asymptotic
approximations for analytical work, such as $\sigma \sim
2\sqrt{\Pe/\pi} \sin(\theta/2)$ as $\Pe \rightarrow \infty$, which is
the fixed point of the dynamics. In this important limit,
\cite{choiADLA} have found an exact similarity solution,
\begin{equation}
g(w,t) = A_1(t) \sqrt{w(w-1)}, \ \ \ A_1(t) = t^{2/3}   \label{eq:adfp}
\end{equation}
to Eq.~(\ref{eq:genPB}) with $\sigma(e^{i\theta},t) = \sqrt{A_1(t)}
\sin(\theta/2)$  (since $\Pe(t) \propto A_1(t)$ for a fixed background
flow). As shown in Figure ~\ref{fig:ADLAcontinuum}, this corresponds
to a constant shape, whose linear size grows like $t^{2/3}$, with a
$90^\circ$ cusp at the rear stagnation point, where a branch point of
$\sqrt{w(w-1)}$ lies on the unit circle.  For any finite, $\Pe(t)$,
however, the cusp is smoothed, and the map remains univalent, although
other singularities may form. Curiously, when mapped to the channel
geometry with $\log z$, the solution (\ref{eq:adfp}) becomes a
Saffman-Taylor finger of width, $\lambda=3/4$.

\section{Stochastic interfacial dynamics}


The continuous dynamics of conformal maps is a mature subject, but
much attention is now focusing on analogous problems with discrete,
stochastic dynamics.  The essential change is in the kinematic
condition: The expression for the interfacial velocity, e.g.
Eq.~(\ref{eq:kinematic}), is re-interpretted as the probability
density (per unit arc length) for advancing the interface with a
discrete ``bump'', e.g. to model a depositing particle.  Continuous
conformal-map dynamics is then replaced by rules for constructing and
composing the bumps.  This method of {\it iterated conformal maps} was
introduced by \cite{hastings98} in the context of Laplacian growth.

\begin{figure}[t]
\begin{center}
\mbox{
(a) \includegraphics[width=0.45\linewidth]{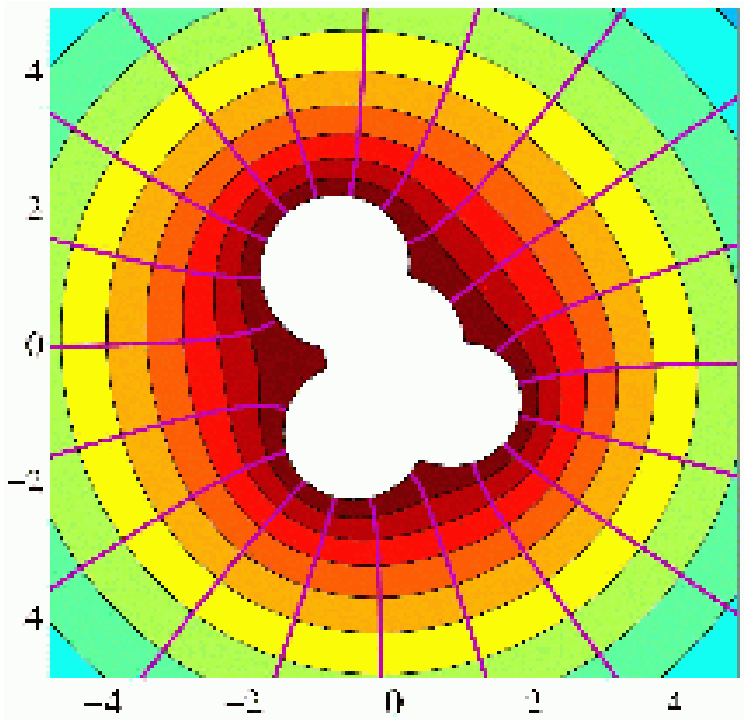}
(b) \includegraphics[width=0.45\linewidth]{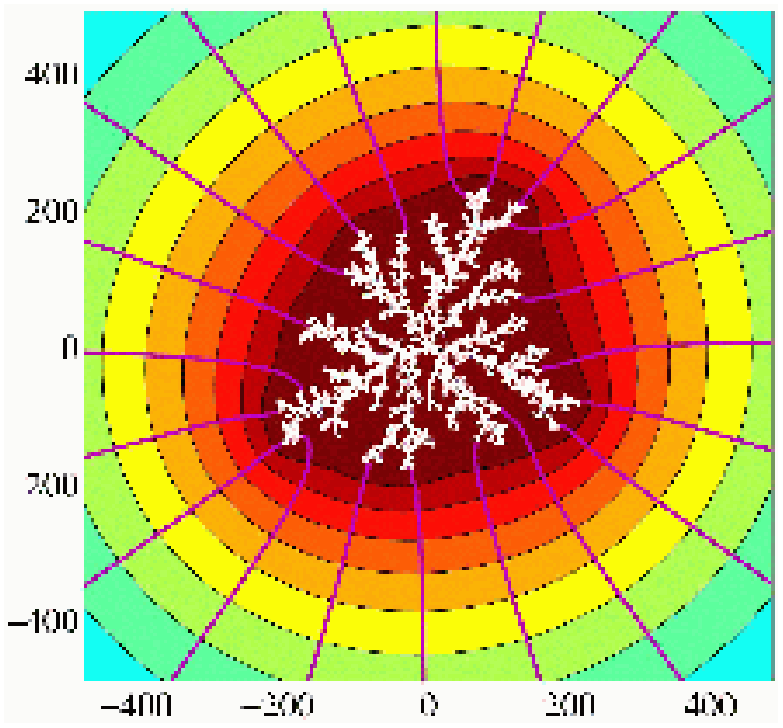}
}
\caption{ \label{fig:DLA}
Simulation of the aggregation of (a) 4 and (b) 10,000 particles using
the Hastings-Levitov algorithm ($a=1/2$). Color contours show the
quasi-steady concentration (or probability) field for mobile particles
arriving from infinity, and purple curves indicate lines of diffusive
flux (or probability current). [Courtesy of Jaehyuk Choi and Benny
Davidovitch.]}
\end{center}
\end{figure}

Stochastic Laplacian growth has been discussed since the early 1980s,
but \cite{hastings98} first showed how to implement it with conformal
mapping. They proposed the following family of {\it bump functions},
\begin{eqnarray}
f_{\lambda,\theta}(w) &=& e^{i\theta} f_\lambda\left( e^{-i\theta}
w\right), \ \ \ |w|\geq 1 \\ f_\lambda(w) &=& w^{1-a}\left[
\frac{(1+\lambda)(w+1)}{2w}\left(
w+1+\sqrt{w^2+1-2w\frac{1-\lambda}{1+\lambda}}\right)-1\right]^a
\label{eq:HLbump}
\end{eqnarray}
as elementary univalent mappings of the exterior of the unit disk used
to advance the interface ($0<a\leq 1$). The function,
$f_{\lambda,\theta}(w)$, places a bump of (approximate) area,
$\lambda$, on the unit circle, centered at angle, $\theta$.  Compared
to analytic functions of the unit disk, the Hastings-Levitov function
(\ref{eq:HLbump}) generates a much more localized perturbation,
focused on the region between two branch points, leaving the rest of
the unit circle unaltered (\cite{benny99}). For $a=1$, the map
produces a {\it strike}, which is a line segment of length
$\sqrt{\lambda}$ emanating normal to the circle. For $a=1/2$, the map
is an invertible composition of simple linear, M\"obius and 
Joukowski transformations, which inserts a semi-circular {\it bump} on
the unit circle. As shown in Figure~\ref{fig:DLA}, this 
yields a good description of aggregating particles,
although other choices, like $a=2/3$, have also been considered
(\cite{benny99}).  Quantifying the effect of the bump shape remains a
basic open question.

Once the bump function is chosen, the conformal map, $z=g_n(w)$, from
the exterior of the unit disk to the evolving domain with $n$ bumps is
constructed by iteration,
\begin{equation}
g_{n}(w) = g_{n-1}\left(f_{\lambda_n,\theta_n}(w)\right)
\end{equation}
starting from the initial interface, given by $g_0(w)$.  All of the
physics is contained in the sequence of bump parameters,
$\{(\lambda_n,\theta_n)\}$, which can be generated in different ways
(in the $w$ plane) to model a variety of physical processes (in the
$z$ plane). As shown in Figure~\ref{fig:DLA}(b), the interface often
develops a very complicated, fractal structure, which is given,
quite remarkably, by an exact mapping of the unit circle.

The great advantage of stochastic conformal mapping over atomistic or
Monte Carlo simulation of interfacial growth lies in its mathematical
insight.  For example, given the sequence $\{(\lambda_n,\theta_n)\}$
from a simulation of some physical growth process, the Laurent
coefficients, $A_k(n)$, of the conformal map, $g_n(w)$, as defined in
Eq.~(\ref{eq:laurent}), can be calculated analytically.  For the bump
function (\ref{eq:HLbump}), \cite{benny99} provide a hierarchy of
recursion relations, yielding formulae such as
\begin{equation}
A_1(n) = \prod_{m=1}^n (1+\lambda_m)^a,  \label{eq:A1}
\end{equation}
and explain how to interpret the Laurent coefficients. For example,
$A_1$ is the {\it conformal radius} of the cluster, a convenient
measure of its linear extent. It is also the radius of a grounded disk
with the same capacitance (with respect to infinity) as the
cluster. The Koebe ``$1/4$ theorem'' on univalent functions
(\cite{duren}) ensures that the cluster (image of the unit disk) is
always contained in a disk of radius $4 A_1$. The next Laurent
coefficient, $A_0$, is the center of a uniformly charged disk, which
would have the same asymptotic electric field as the cluster (if also
charged). Similarly, higher Laurent coefficients encode higher
multipole moments of the cluster.

Mapping the unit circle with a truncated Laurent expansion defines the
{\it web}, which wraps around the growing tips and exhibits a sort of
turbulent dynamics, endlessly forming and smoothing cusp-like
protrusions (\cite{hastings97,hastings98}). The stochastic dynamics,
however, does not suffer from finite-time singularities because the
iterated map, by construction, remains univalent.  In some sense,
discreteness plays the role of surface tension, as an another
regularization of ill-posed continuum models like Laplacian growth.

\noindent
{\bf{Diffusion-Limited Aggregation (DLA):}} The stochastic analog of
Laplacian growth is the DLA model of \cite{witten81}, illustrated in
Figure~\ref{fig:DLA}, in which particles perform random walks
one-by-one from inifinity until they stick irreversibly to a cluster,
which grows from a seed at the origin.  DLA and its variants (see
below) provide simple models for many fractal patterns in nature, such
as colloidal aggregates, dendritic electro-deposits, snowflakes,
lightning strikes, mineral deposits, and surface patterns in ion-beam
microscopy (\cite{bunde}). In spite of decades of research, however,
DLA still presents theoretical mysteries, which are just beginning to
unravel (\cite{halsey00}).

The Hastings-Levitov algorithm for DLA prescribes the bump parameters,
$\{(\lambda_n,\theta_n)\}$, as follows. As in Laplacian growth, the
harmonic function for the concentration (or probability density) of
the next random walker approaching an $n$-particle cluster is simply,
\begin{equation}
\phi_n(z) = A \, \Real \log g^{-1}_n(z),   \label{eq:log}
\end{equation}
according to Eq.~(\ref{eq:phitrans}), since $\phi(w) = A\, \Real \log w =
\log |w|$ is the (trivial) solution to Laplace's equation in the
mathematical $w$ plane with $\phi=0$ on the unit disk with a
circularly symmetric flux density, $A$, prescribed at infinity. Using
the transformation rule, Eq.~(\ref{eq:gradtrans}), we then find that
the evolving {\it harmonic measure}, $p_n(z)|dz|$, for the $n$th
growth event corresponds to a uniform probability measure,
$P_n(\theta)d\theta$, for angles, $\theta_n$, on the unit circle,
$w=e^{i\theta}$:
\begin{equation}
p_n(z)|dz| = |\nabla_z \phi| |dz| = \left| \frac{\nabla_w
\phi}{\overline{g^\prime_{n-1}}}\right| |g_{n-1}^\prime \, dw| = |\nabla_w \phi|
|dw| = \frac{d\theta}{2\pi}= P_n(\theta)d\theta,    \label{eq:measure}
\end{equation}
where we set $A=1/2\pi$ for normalization, which implicitly sets the
time scale (see below).  The conformal invariance of the harmonic
measure is well known in mathematics, but the surprising result of
\cite{hastings98} is that all the complexity of DLA is slaved to a
sequence of independent, uniform random variables.

Where the complexity resides is in the bump area, $\lambda_n$, which
depends non-trivially on current cluster geometry and thus on the
entire history of random angles, $\{\theta_m | m\leq n\}$. For DLA,
the bump area in the mathematical $w$ plane should be chosen such that
it has a fixed value, $\lambda_0$, in the physical $z$ plane, equal to
the aggregating particle area. As long as the new bump is sufficiently
small, it is natural to try to correct only for the Jacobian factor
\begin{equation}
 J_{n}(w) = |g^\prime_{n}(w)|^2  = \prod_{m=1}^{n}
|g^\prime_{\lambda_m,\theta_m}(w)|^2 \label{eq:J}
\end{equation}
of the previous conformal map at the center of the new bump,
\begin{equation}
\lambda_n = \frac{\lambda_0}{J_{n-1}(e^{i\theta_n})},   \label{eq:lambda}
\end{equation}
although it is not clear {\it a priori} that such a local
approximation is valid. Note at least that $g^\prime_n \rightarrow
\infty$, and thus $\lambda_n \rightarrow 0$, as the cluster grows, so
this has a chance of working.

\begin{figure}
\begin{center}
\mbox{
(a) \includegraphics[width=0.45\linewidth]{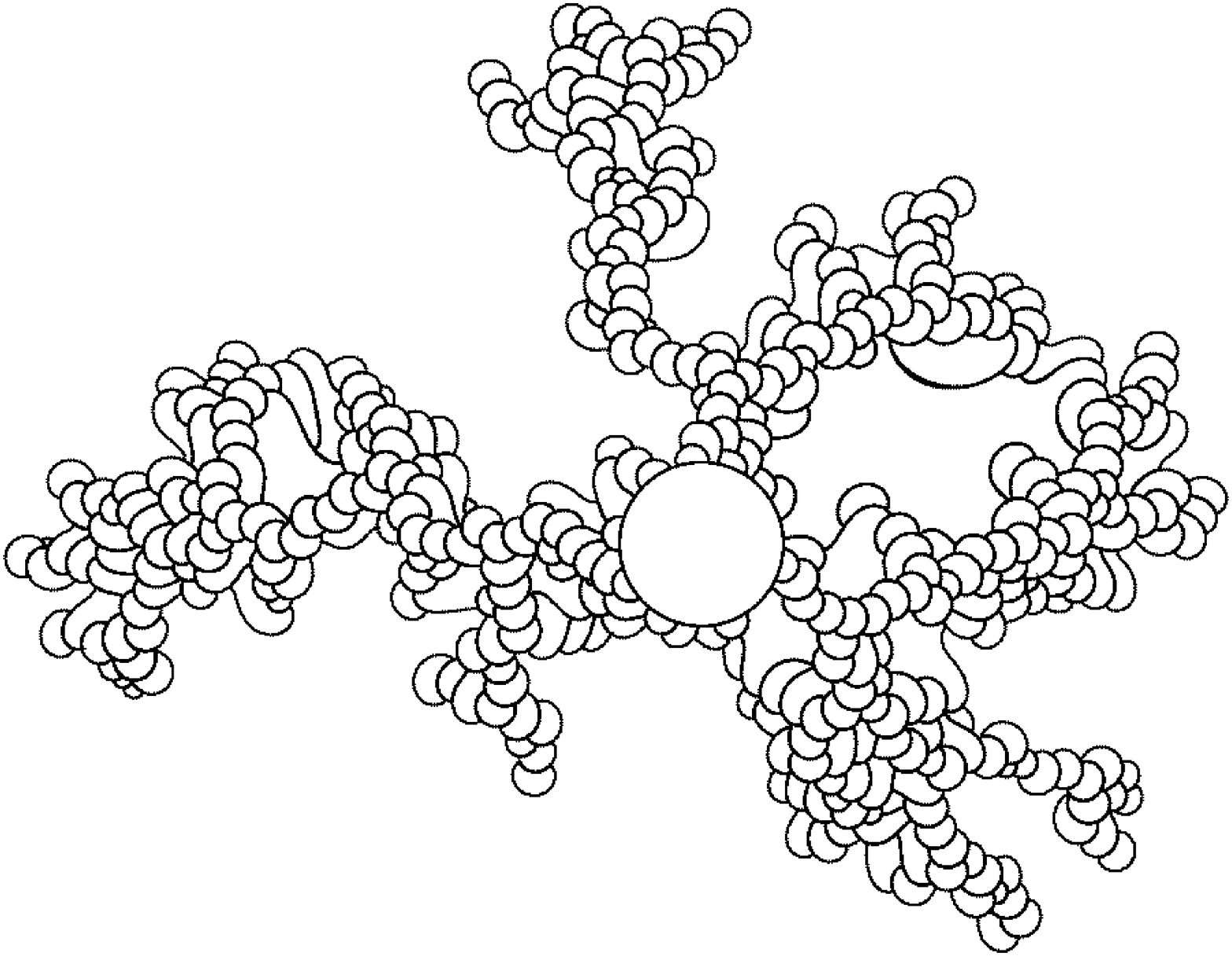} \nolinebreak
(b) \includegraphics[width=0.45\linewidth]{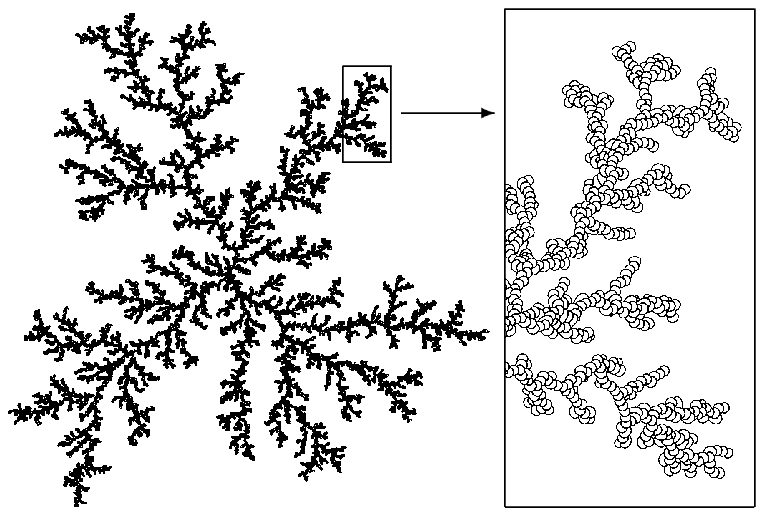}
}\\
\mbox{
(c) \includegraphics[width=0.45\linewidth]{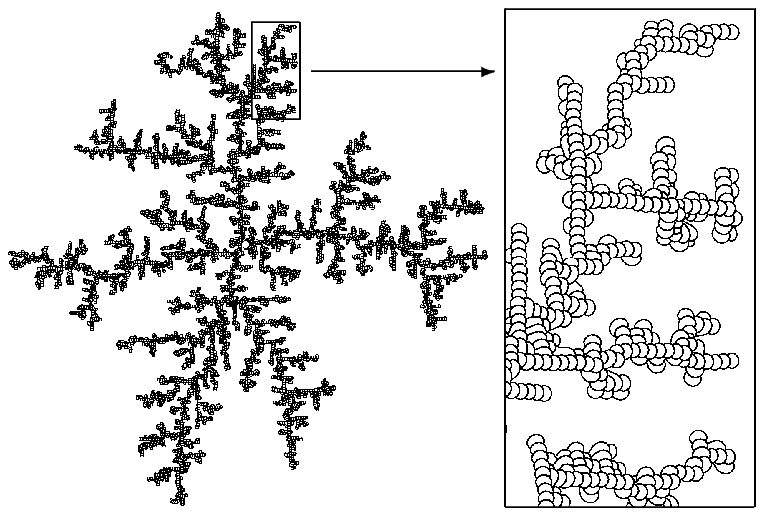}
(d) \includegraphics[width=0.45\linewidth]{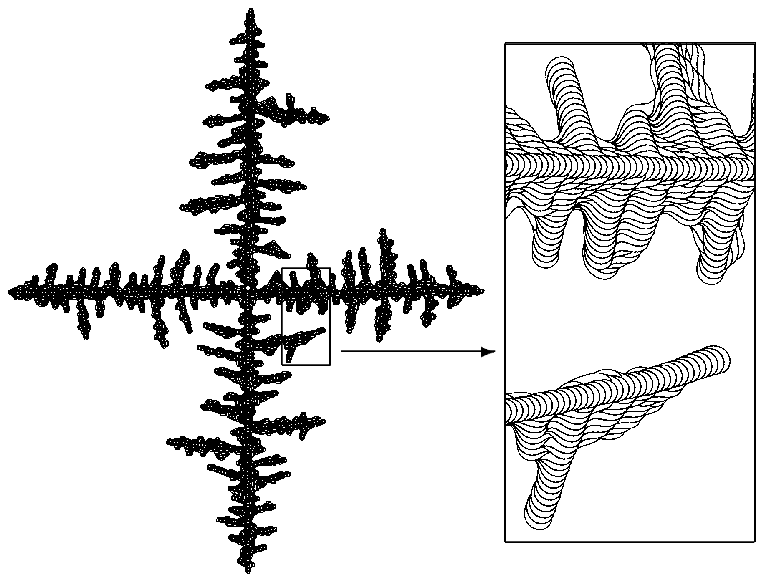}
} \\
\mbox{
(e) \includegraphics[width=0.45\linewidth]{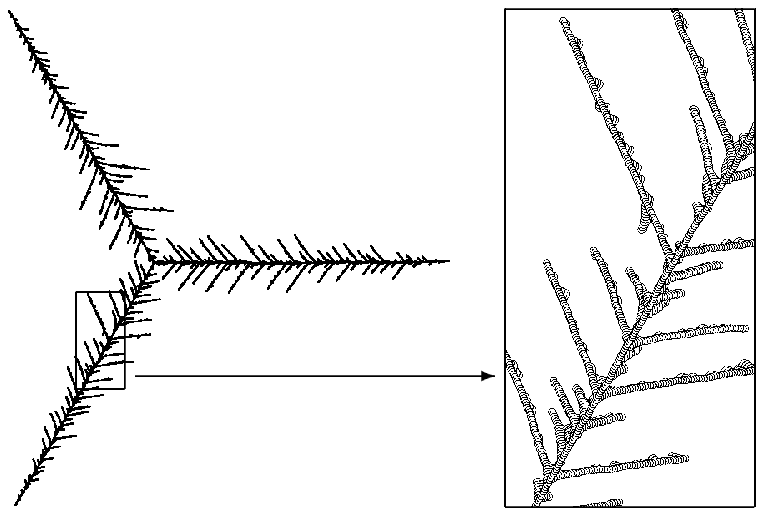} \nolinebreak
\nolinebreak
(f) \includegraphics[width=0.45\linewidth]{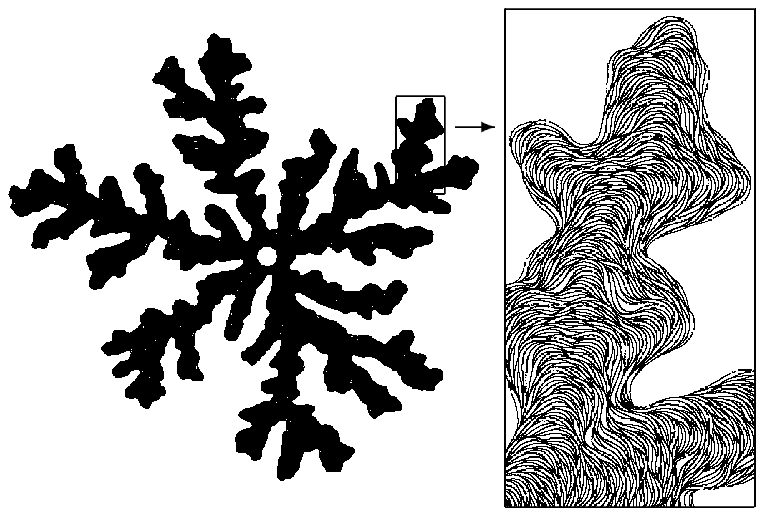}
}
\caption{ \label{fig:SL}
Simulations of fractal aggregates by ~\cite{stepanov01}: (a)
Superimposed time series of the boundary, showing the aggregation of
particles, represented by iterated conformal maps; (b) a larger
simulation with a particle-area acceptance window; (c) the result of
anisotropic growth probability with square symmetry; (d)
square-anisotropic growth with noise control via flat particles; (e)
triangular-anisotropic growth with noise control; (c) isotropic growth
with noise control, which resembles radial viscous
fingering. [Courtesy of Leonid Levitov.]  }
\end{center}
\end{figure}

Numerical simulations with the Hastings-Levitov algorithm do indeed
produce nearly constant bump areas, as in
Figure~\ref{fig:DLA}. Nevertheless, much larger ``particles'', which
fill deep fjords in the cluster, occasionally occur where the map
varies too wildly, as shown in Figure~\ref{fig:SL}(a). It is possible
(but somewhat unsatisfying) to reject particles outside an ``area
acceptance window'' to produce rather realistic DLA clusters, as shown
in Figure~\ref{fig:SL}(b). It seems that the rejected large bumps are
so rare that they do not much influence statistical scaling properties
of the clusters (\cite{stepanov01}), although this issue is by no
means rigorously resolved.

\noindent
{\bf Fractal geometry:} Fractal patterns abound in nature, and DLA
provides the most common way to understand them \cite{bunde}. The
fractal scaling of DLA has been debated for decades, but conformal
dynamics is shedding new light on the problem. Simulations show that
the conformal radius (\ref{eq:A1}) exhibits fractal scaling, $A_1(n)
\propto n^{1/D_f}$, where the {\it fractal dimension}, $D_f=1.71$, agrees
with the accepted value from Monte Carlo (random walk) simulations of
DLA, although the prefactor seems to depend on the bump function
(\cite{benny99}). A perturbative renormalization-group analysis of the
conformal dynamics by \cite{hastings97} gives a similar result, $D_f =
2-1/2+1/5 = 1.7$.  The multifractal spectrum of the harmonic measure
has also been studied (\cite{jensen02,ball02}).

Perhaps the most basic question is whether DLA clusters are truly
fractal -- statistically self-similar and free of any length
scale. This long-standing question requires accurate statistics and
very large simulations, to erase the surprisingly long memory of the
initial conditions. Conformal dynamics provides exact formulae for
cluster moments, but simulations are limited to at most $10^5$
particles by poor $O(n^2)$ scaling, caused by the history-dependent
Jacobian in Eq.~(\ref{eq:lambda}). In contrast, efficient random-walk
simulations can aggregate many millions of particles.

Therefore, \cite{somfai99} developed a hybrid method relying only upon
the existence of the conformal map, but not the Hastings-Levitov
algorithm to construct it.  Large clusters by Monte Carlo simulation,
and approximate Laurent coefficients are computed, purely for their
morphological information, as follows. For a given cluster of size
$N$, $M$ random walkers are launched from far away, and the positions,
$z_m$, where they would first touch the cluster, are recorded. If the
conformal map, $z=g_n(e^{i\theta})$, were known, the points $z_m$
would correspond to $M$ angles $\theta_m$ on the unit circle. Since
these must sample a uniform distribution, one assumes $\theta_m = 2\pi
m/M$ for large $M$. From Eq.~(\ref{eq:laurent}), the Laurent
coefficients are simply the Fourier coefficients of the discretely
sampled function, $z_m = \sum A_k e^{i\theta_m k}$. Using this method,
all Laurent coefficients appear to scale with the same fractal
dimension,
\begin{equation}
\langle |A_k(n)|^2 \rangle \propto n^{2/D_f}
\end{equation}
although the first few coefficients crossover extremely slowly to the
asymptotic scaling. 

\noindent
{\bf{Snowflakes and viscous fingers:}} In conventional Monte Carlo
simulations, many variants of DLA have been proposed to model real
patterns found in nature (\cite{bunde}).  For example, clusters
closely ressembling snowflakes can be grown by a combination of noise
control (requiring multiple hits before attachment) and anisotropy (on
a lattice).  Conformal dynamics offers the same flexibility, as shown
in Figure~\ref{fig:SL}, while allowing anisotropy and noise to be
controlled independently (\cite{stepanov01}). Anisotropy can be
introduced in the growth probability with a weight factor, $1+c \cos
m\alpha_n$, where $\alpha_n$ is the surface orientation angle in the
physical plane given by Eq.~(\ref{eq:aniso}), or by simply rejecting
angles outside some tolerance from the desired symmetry
directions. Noise can be controlled by flattening the aspect ratio of
the bumps. Without anisotropy, this produces smooth fluid-like
patterns (Figure~\ref{fig:SL}(f)), reminiscent of viscous fingers
(Figure~\ref{fig:siegel}).

The possible relation between DLA and viscous fingering is a
tantalizing open question in pattern formation. Many authors have
argued that the regularization of finite-time singularities in
Laplacian growth by discreteness is somehow analogous to surface
tension.  Indeed, the average DLA cluster in a channel, grown by
conformal mapping, is similar (but not identical) to a Saffman-Taylor
finger of width 1/2 (\cite{somfai03}), and the instantaneous expected
growth rate of a cluster can be related to the Polubarinova-Galin (or
``Shraiman-Bensimon'') equation (\cite{hastings98}).  Conformal
dynamics with many bumps grown simultaneously suggests that Laplacian
growth and DLA are in different universality classes, due to the basic
difference of layer-by-layer versus one-by-one growth, respectively
(\cite{benny02}). Another multiple-bump algorithm with complete
surface coverage, however, seems to yield the opposite conclusion
(\cite{levermann04}).

\begin{figure}
\begin{center}
\mbox{
(a) \includegraphics[width=0.41\linewidth]{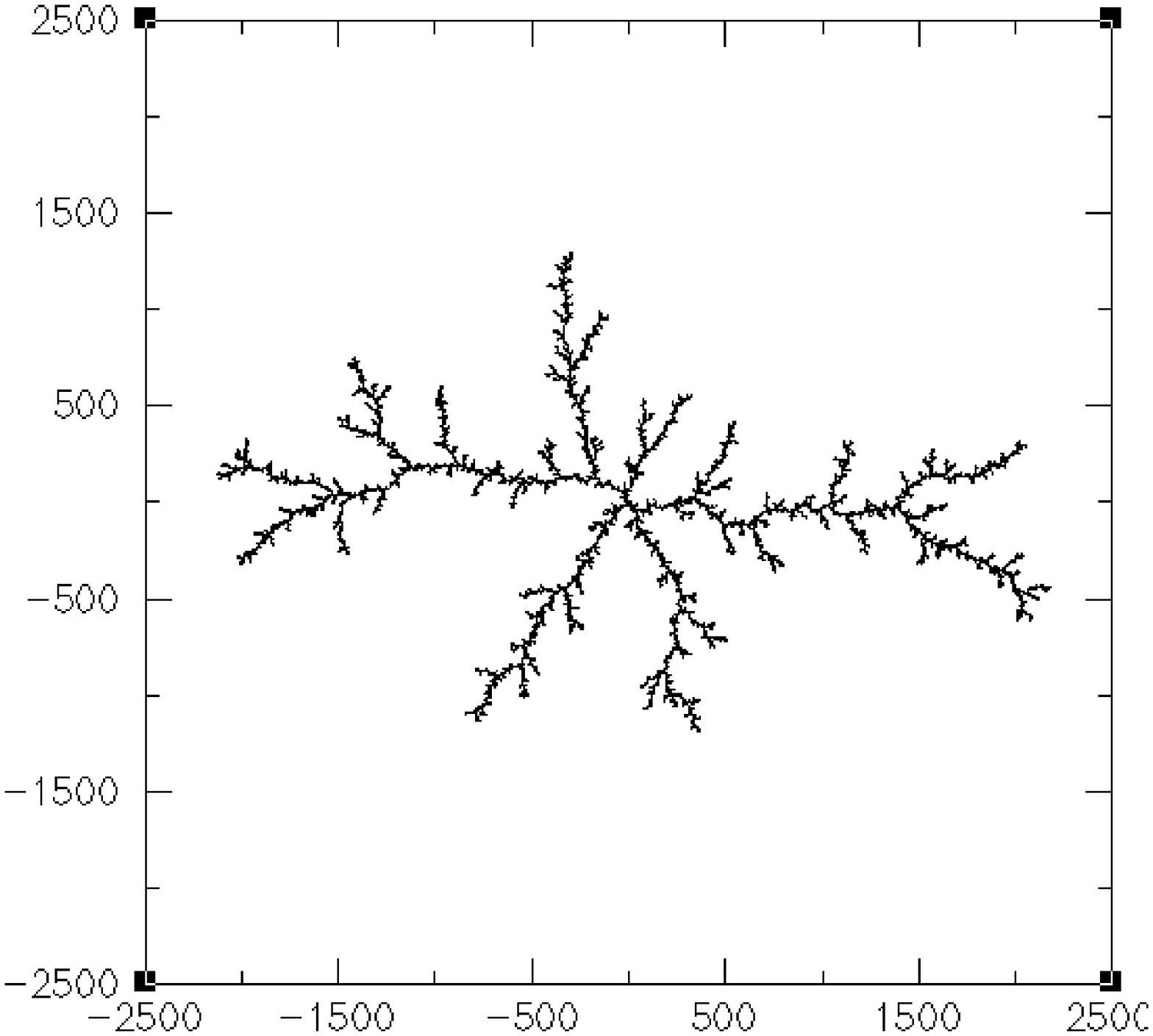}
\nolinebreak
(b) \includegraphics[width=0.4\linewidth]{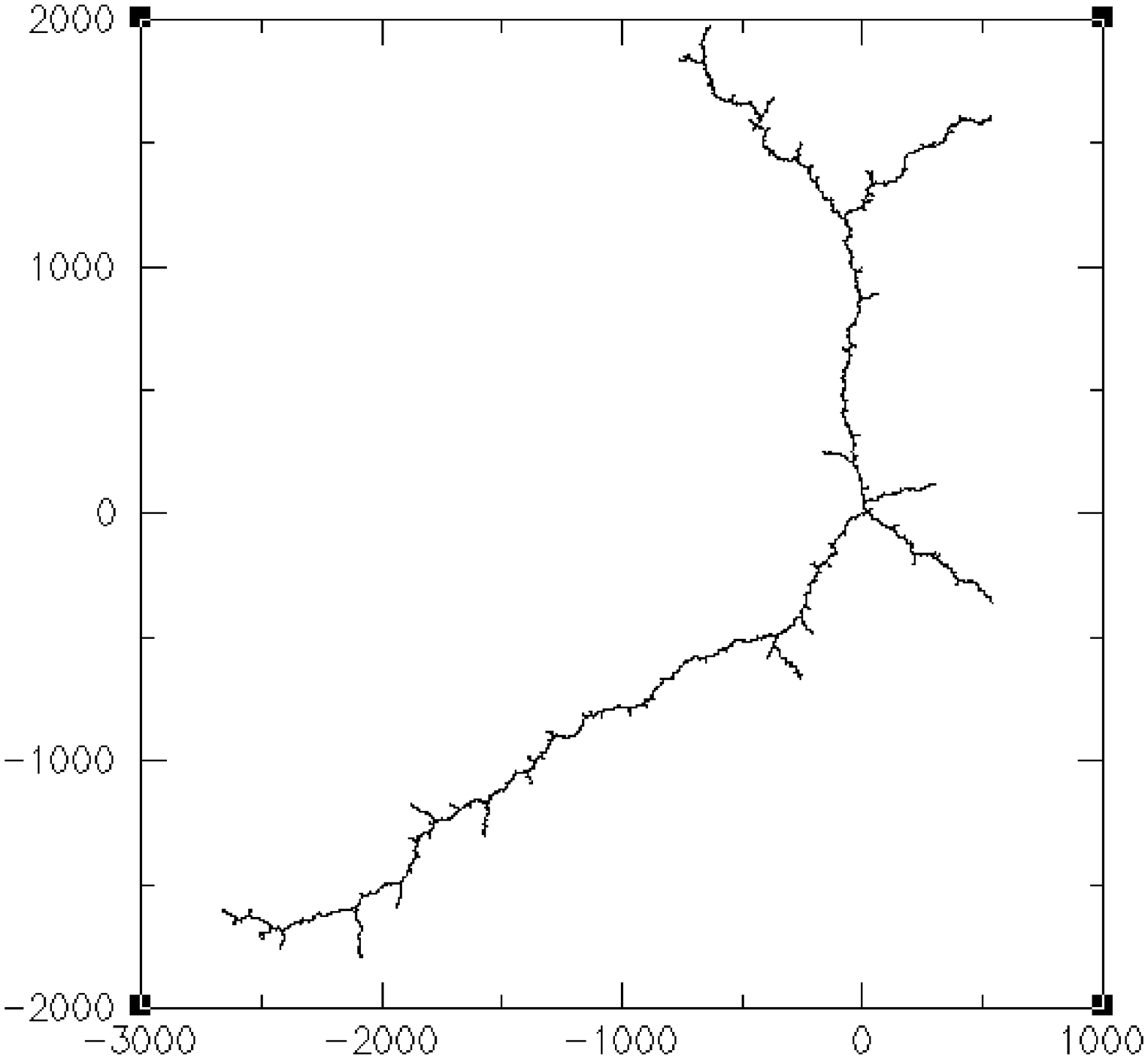}
}
\caption{ \label{fig:DBM} Conformal-mapping simulations by \cite{hastings01} of the
Dielectric Breakdown Model with (a) $\eta=2$ and (b) $\eta=3.5$.
[Courtesy of Matt Hastings.] }
\end{center}
\end{figure}

\noindent 
{\bf{Dielectric breakdown:}} In their original paper,
\cite{hastings98} allowed for the size of the bump in the physical
plane to vary with an exponent, $\alpha$, by replacing $J_{n-1}$ with
$(J_{n-1})^{\alpha/2}$ in Eq.~(\ref{eq:lambda}). In DLA ($\alpha=2$),
the bump size is roughly constant, but for $0<\alpha<2$ the bump size
grows with the local gradient of the Laplacian field. This is a simple
model for dielectric breakdown, where the stochastic growth of an
electric discharge penetrating a material is nonlinearly enhanced by
the local electric field.  One could use strikes ($a=0$) rather than
bumps ($a=1/2$) to better reproduce the string-like branched patterns
seen in laboratory experiments (\cite{bunde}) and more familiar
lightning strikes. The model displays a ``stable-to-turbulent'' phase
transition: The relative surface roughness decreases with time for
$0\leq \alpha < 1$ and grows for $\alpha>1$.

The original Dielectric Breakdown Model (DBM) of \cite{niemeyer84} has
a more complicated conformal-dynamics representation. As usual, the
growth is driven by the gradient of a harmonic function, $\phi$, (the
electrostatic potential) on an iso-potential surface (the discharge
region). Unlike the $\alpha$-model above, however, DBM growth events
are assumed to have constant size, so the bump size in the
mathematical plane is still chosen according to
Eq.~(\ref{eq:lambda}). The difference lies in the growth measure,
which does not obey Eq.~(\ref{eq:measure}). Instead, the generalized
harmonic measure in the physical $z$ plane is given by
\begin{equation}
p(z) \propto |\nabla_z \phi|^\eta
\end{equation}
where $\eta$ is an exponent interpolating between the Eden model
($\eta=0$), DLA ($\eta=1$), and nonlinear dielectric breakdown
($\eta>1$).  For $\eta\neq 1$, the fortuitous cancellation in
Eq.~(\ref{eq:measure}) does not occur. Instead, a similar calculation
using Eq.~(\ref{eq:log}) yields a non-uniform probability measure for
the $n$th angle on the unit circle in the mathematical plane,
\begin{equation}
P_n(\theta_n) = |g_{n-1}^\prime(e^{i\theta_n})|^{1-\eta} 
= \prod_{m=1}^{n-1} |f_{\lambda_m,\theta_m}(e^{i\theta_n})|^{1-\eta}
\end{equation}
which is complicated and depends on the entire history of the
simulation.

Nevertheless, conformal mapping can be applied fruitfully to DBM,
because not solving Laplace's equation around the cluster outweighs
the difficulty of sampling the angle measure. Surmouting the latter
with a Monte Carlo algorithm, \cite{hastings01} has performed DBM
simulations of $10^4$ growth events, an order of magnitude beyond
standard methods solving Laplace's equation on a lattice. The results,
illustrated in Figure~\ref{fig:DBM}, support the theoretical
conjecture that DBM clusters become one-dimensional, and thus
non-fractal, for $\eta
\geq 4$.

Using the conformal-mapping formalism, efforts are also underway to
develop a unified scaling theory of the $\eta$-model for the growth
probability from DBM combined with the $\alpha$-model above for the
bump size (\cite{ball02}).

\noindent
{\bf{Brittle fracture:}} Modeling the stochastic dynamics of fracture
is a daunting problem, especially in heterogeneous materials
(\cite{bunde,hermann}). The basic equations and boundary conditions
are still the subject of debate, and even the simplest models are
difficult to solve.  In two dimensions, stochastic conformal mapping
provides an elegant, new alternative to discrete-lattice and
finite-element models.

In brittle fracture, the bulk material is assumed to obey Lam\'e's
equation of linear elasticity,
\begin{equation}
\rho \frac{\partial^2 \ub}{\partial t^2} =
(\lambda+\mu)\del(\del\cdot\ub) + \mu \del^2\ub    \label{eq:lame}
\end{equation}
where $\ub$ is the displacement field, $\rho$ is the density, and
$\mu$ and $\lambda$ are Lam\'e's constants. For conformal mapping, it
is crucial to assume (i) two-dimensional symmetry of the fracture
pattern and (ii) quasi-steady elasticity, which sets the left hand
side to zero to obtain equations of the type described above. For Mode
III fracture, where a constant out-of-plane shear stress is applied at
infinity, we have $\del\cdot\ub=0$, so the steady Lam\'e equation
reduces to Laplace's equation for the out-of-plane displacement,
$\del^2 u_z=0$, which allows the use of complex potentials. For Modes
I and II, where a uniaxial, in-plane tensile stress is applied at
infinity, the steady Lam\'e equation must be solved. As discussed
above, this is equivalent to the bi-harmonic equation for the Airy
stress function, which allows the use of Goursat functions.

For all three modes, the method of iterated conformal maps can be
adapted to produce fracture patterns for a variety of physical
assumptions about crack dynamics (\cite{barra02}). For Modes I and II
fracture, these models provide the first examples of stochastic
bi-harmonic growth, which have interesting differences with stochastic
Laplacian growth for Mode III fracture.  The Hastings-Levitov
formalism is used with constant-size bumps, as in DLA, to represent
the fracture process zone, where elasticity does not apply. The growth
measure a function of the excess tangential stress, beyond a critical
yield stress, $\sigma_c$, characterizing the local strength of the
material. Quenched disorder is easily included by making $\sigma_c$ a
random variable.  In spite of its many assumptions, the method
provides analytical insights, while obviating the need to solve
Eq.~(\ref{eq:lame}) during fracture dynamics, so it merits futher
study.

\noindent 
{\bf{Advection-Diffusion-Limited Aggregation:}} Non-local fractal
growth models typically involve a single bulk field driving the
dynamics, such as the particle concentration in DLA, the electric
field in DBM, or the strain field in brittle fracture, and as a result
these models tend to yield statistically similar structures, apart
from the effect of boundary conditions. Pattern formation in nature,
however, is often fueled by multiple transport processes,
such as diffusion, electromigration, and/or advection in a fluid
flow. The effect of such dynamical competition on growth morphology is
an open question, which would be difficult to address with
lattice-based or finite-element methods, since many large fractal
clusters must be generated to fully explore the space and time dependence.

Once again, conformal mapping provides a convenient means to formulate
stochastic analogs of the non-Laplacian transport-limited growth
models from \S 2.3 (in two dimensions). It is straightforward to adapt
the Hastings-Levitov algorithm to construct stochastic dynamics driven
by bulk fields satsifying the conformally invariant system of
equations~(\ref{eq:geneq}). A class of such models has recently been
formulated by \cite{ADLA}.

\begin{figure}
\begin{center}
\mbox{
\includegraphics[width=0.45\linewidth]{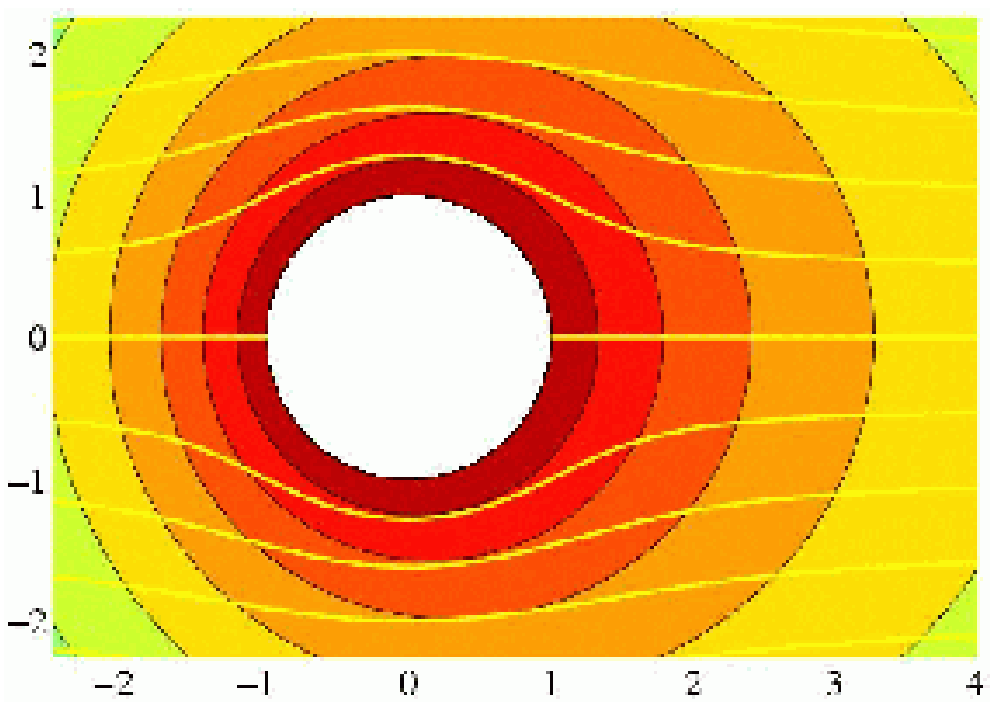} \nolinebreak
\includegraphics[width=0.45\linewidth]{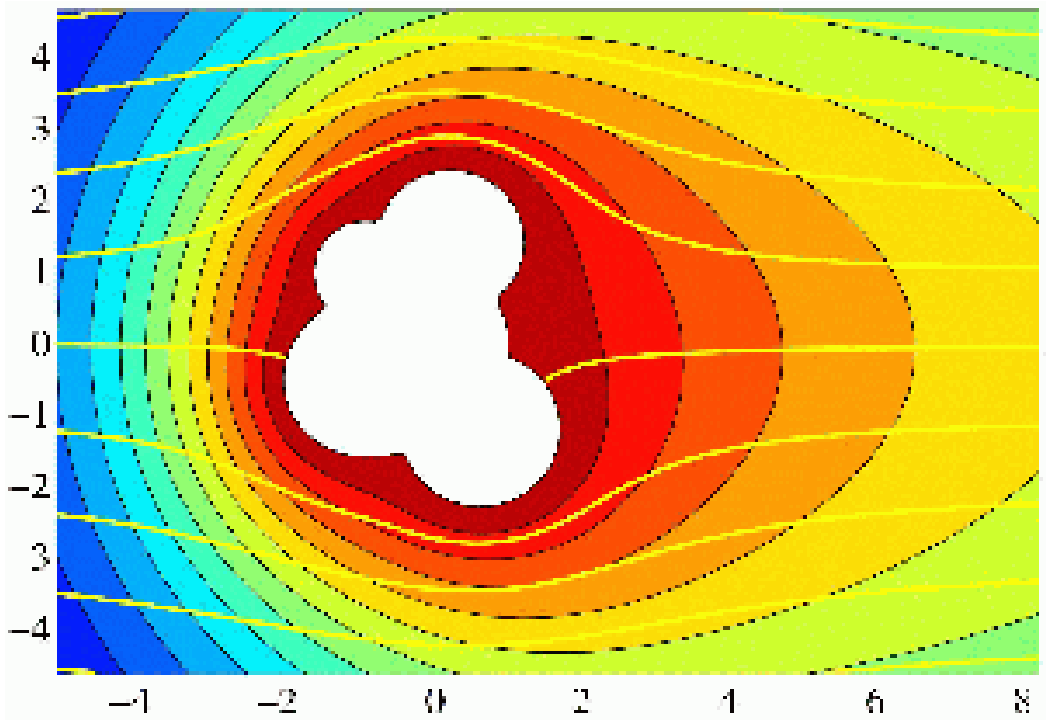}} \\
\mbox{
\includegraphics[width=0.45\linewidth]{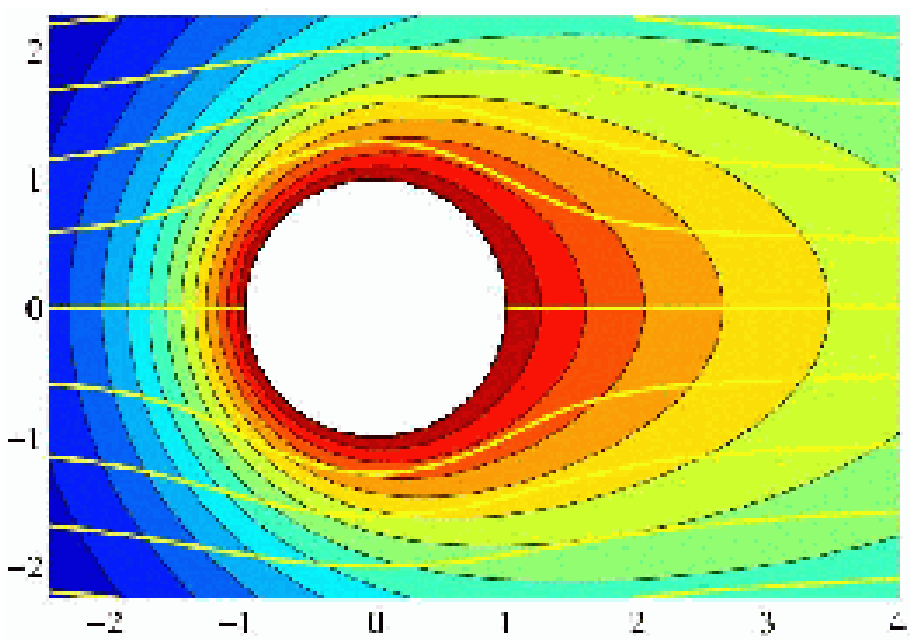} \nolinebreak
\includegraphics[width=0.45\linewidth]{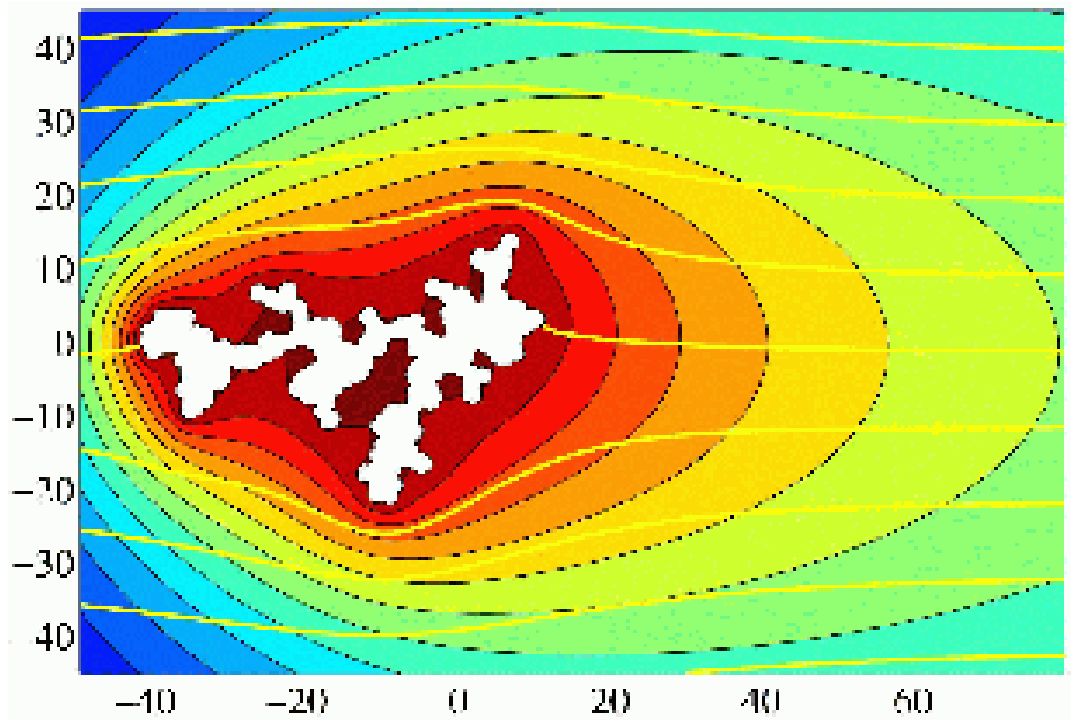}} \\
\mbox{
\includegraphics[width=0.45\linewidth]{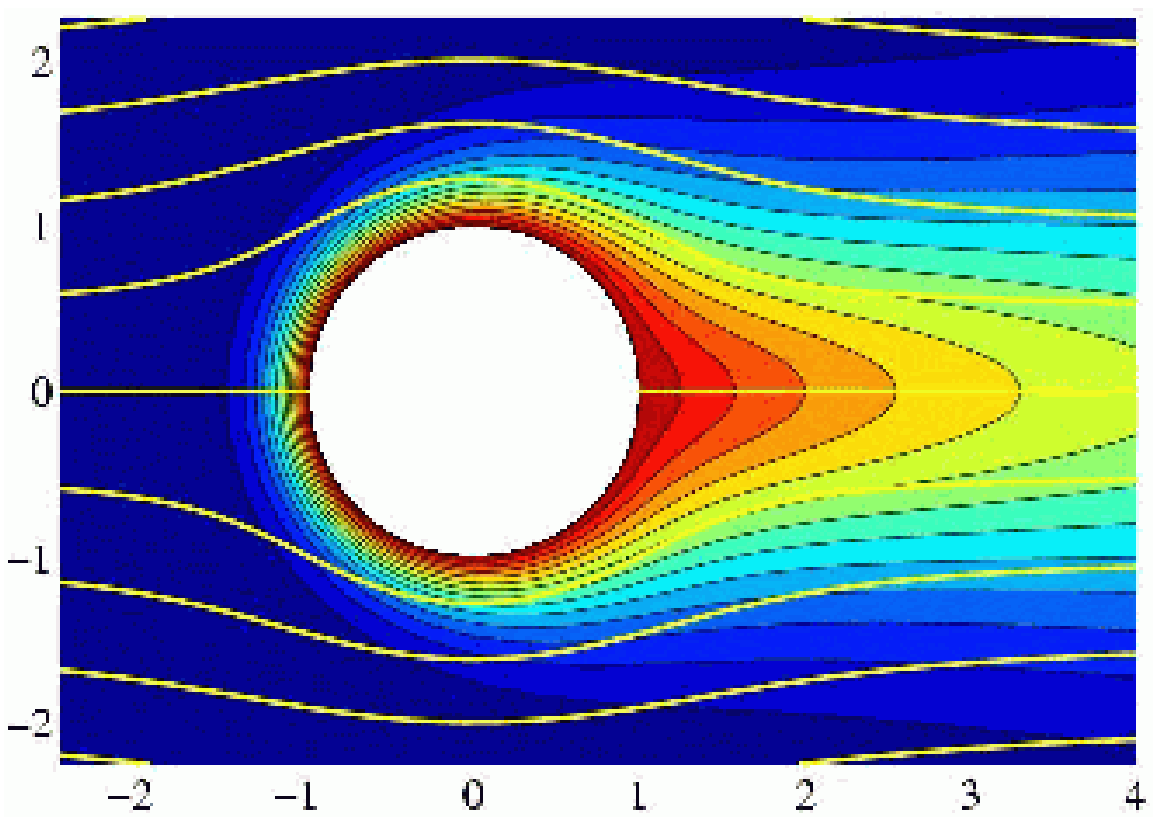} \nolinebreak
\includegraphics[width=0.45\linewidth]{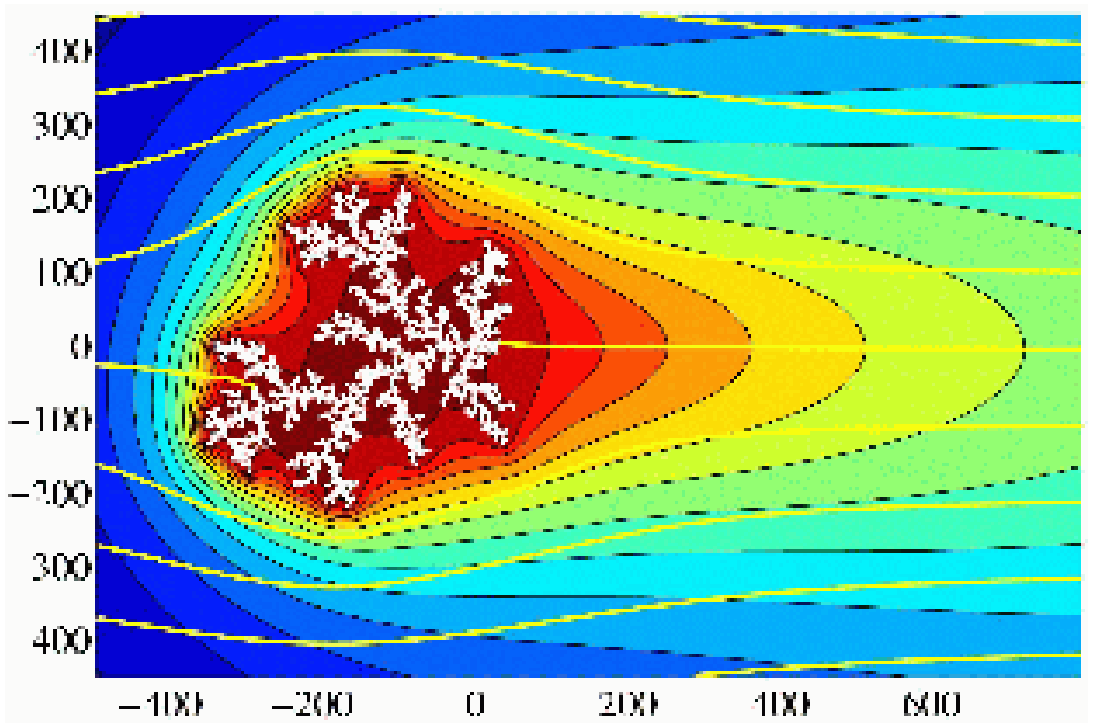} 
}
\caption{ \label{fig:ADLA} A simulation of Advection-Diffusion-Limited
Aggregation from ~\cite{ADLA}. In each row, the growth probabilities in
the physical $z$-plane (on the right) are obtained by solving
advection-diffusion in a potential flow past an absorbing cylinder in
the mathematical $w$-plane (on the left), with the same time-dependent
P\'elcet number.}
\end{center}
\end{figure}

Perhaps the simplest case involving two transport processes,
illustrated in Figure~\ref{fig:ADLA}, is Advection-Diffusion-Limited
Aggregation (ADLA), or ``DLA in a flow''. Imagine a fluid carrying a
dilute concentration of sticky particles flowing past a sticky object,
which begins to collect a fractal aggregate. As the cluster grows, it
causes the fluid to flow around it and changes the concentration
field, which in turn alters the growth probability measure.  Assuming
a quasi-steady potential flow with a uniform speed far from the
cluster, the dimensionless transport problem is
\begin{eqnarray}
\Pe_0 \del \phi \cdot \del c = \del^2 c , \ \ \del^2 \phi = 0 ,  \ & &
  z \in \Omega_z(t)  \\ 
c=0 , \ \ \nhat\cdot\del\phi=0 ,
 \ \ \sigma = \nhat\cdot\del c, \  & &  z \in \partial\Omega_z(t) \\ 
c \to 1 , \ \ \del\phi
 \to \xhat, \  & &  |z| \to \infty 
\end{eqnarray}
where $\Pe_0$ is the initial P\'eclet number and $\sigma$ is the
diffusive flux to the surface, which drives the growth. The transport
problem is solved in the mathematical $w$ plane, where it corresponds
to a uniform potential flow of concentrated fluid past an absorbing
circular cylinder. The normal diffusive flux on the cylinder,
$\sigma(\theta,\Pe)$, can be obtained from a tabulated numerical
solution or an accurate analytical approximation (\cite{advdif}).

Because the boundary condition on $\phi$ at infinity is not
conformally invariant, the flow in the $w$ plane has a time-dependent
P\'eclet number, $\Pe(t) = A_1(t) \Pe_0$, which grows with the
conformal radius of the cluster. As a result, the probability of the
$n$th growth event is given by a time-dependent, non-uniform measure
for the angle on the unit circle,
\begin{equation}
P_n(\theta_n ) = \frac{\beta}{\lambda_0} \, \tau_n \,
\sigma(e^{i\theta_n},A_1(t_{n-1})) ,  
\end{equation}
where $\beta$ is a constant setting the mean growth rate.
The waiting time between
growth events is an exponential random variable with mean, $\tau_n$,
given by the current integrated flux to the object,
\begin{equation}
\frac{\lambda_0}{\beta \tau_n}  =  \int_0^{2\pi}
\sigma(e^{i\theta},A_1(t_{n-1})) \, d\theta .  
\end{equation}
Unlike DLA, the aggregation speeds up as the cluster grows, due to a 
larger cross section to catch new particles in the flow.

As shown in Figure~\ref{fig:ADLA}, the model displays a universal
dynamical crossover from DLA (the unstable fixed point) to an 
advection-dominated stable fixed point, since $\Pe(t) \rightarrow
\infty$.  Remarkably, the fractal dimension remains constant during the
transition, equal to the value for DLA, in spite of dramatic changes
in the growth rate and morphology (as indicated by higher Laurent
coefficients).  Moreover, the shape of the ``average'' ADLA cluster in
the high-$\Pe$ regime of Figure~\ref{fig:ADLA} is quite similar (but
not identical) to the exact solution, Eq.~(\ref{eq:adfp}), for the
analogous continuous problem in Figure~\ref{fig:ADLAcontinuum}. Much
remains to be done to understand these kinds of models
and apply them to materials problems.

\section{Curved Surfaces }

\cite{entov91} considered the generalized problem of
Hele-Shaw flows in a non-planar cell having non-zero curvature. In
such problems, the velocity of the viscous flow is still the (surface)
gradient of a potential, $\phi$, but this function is now a
solution of the so-called {\it Laplace-Beltrami equation} on the curved
surface.  The Riemann mapping theorem extends to curved surfaces and
says that any simply-connected smooth surface is conformally
equivalent to the unit disk, the complex plane, or the Riemann sphere.
A common example is the well-known {\it stereographic projection} of
the surface of a sphere to the (compactified) complex plane.  Under a
conformal mapping, solutions of the Laplace-Beltrami equation map to
solutions to Laplace's equation and this combination of facts led
\cite{entov91} to identify classes of explicit
solutions to the continuous Hele-Shaw problem in a variety of
non-planar cells.  With very similar intent, \cite{parisio01} have
recently considered the evolution of Saffman-Taylor fingers on the
surface of a sphere.

\begin{figure}[t]
\begin{center}
\mbox{
\includegraphics[width=0.5\linewidth]{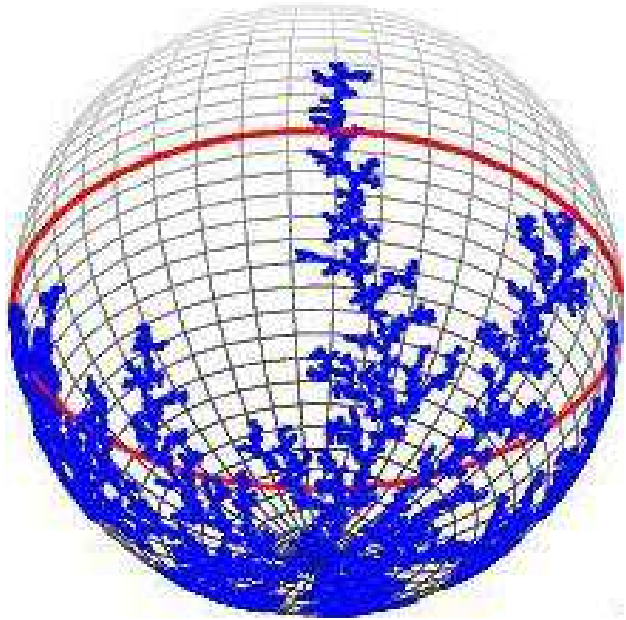}
}
\caption{ \label{fig:DLAsphere} Conformal-mapping simulation of DLA on
a sphere. Particles diffuse one by one from the North Pole and
aggegrate on a seed at a South Pole. [Courtesy of Jaehyuk Choi, Martin
Bazant, and Darren Crowdy.]}
\end{center}
\end{figure}

By now, the reader may realize that most of the methods already
considered in this article are, in principle, amenable to
generalization to curved surfaces, which can be reached by conformal
mapping of the plane.  For example, Figure~\ref{fig:DLAsphere} shows a
simulation of a DLA cluster growing on the surface of a sphere, using
a generalized Hastings-Levitov algorithm, which takes surface curvature
into account.  The key modification is to multiply the Jacobian in
Eq.~(\ref{eq:J}) by the Jacobian of the stereographic projection,
$1+|z/R|^2$, where $R$ is the radius of the sphere.

It should also be clear that any continuous or discrete growth model
driven by a conformally-invariant bulk field, such as ADLA, can be
simulated on general curved surfaces by means of appropriate conformal
projection to a complex plane. The reason is that the system of
equations (\ref{eq:geneq}) is invariant under any conformal mapping,
to a flat or curved surface, because each term transforms like the
Laplacian, $\del^2 \phi \rightarrow J \del^2\phi$, where $J$ is the
Jacobian.  The purpose of studying these models is not only to
understand growth on a particular ideal shape, such as a sphere, but
more generally to explore the effect of local surface curvature on
pattern formation. For example, this could help interpret mineral
deposit patterns in rough geological fracture surfaces, which form by
the diffusion and advection of oxygen in slwoly flowing water.

\section{Outlook}

Although conformal mapping has been with us for centuries, new
developments with applications continue to the present day. This
appears to be the first pedagogical review of stochastic
conformal-mapping methods for interfacial dynamics, which also covers
the latest progress in continuum methods.  Hopefully, this will
encourage the futher exchange of ideas (and people) between the two
fields. Our focus has also been on materials problems, which provide
many opportunites to apply and extend conformal mapping.

Building on specific open questions scattered throughout the text,
we close with a general outlook on directions for future research. A
basic question for both stochastic and continuum methods is the effect
of geometrical constraints, such as walls or curved surfaces, on
interfacial dynamics. Most work to date has been for either radial or
channel geometries, but it would be interesting to describe finite
viscous fingers or DLA clusters growing near walls of various shapes,
as is often the case in materials applications.

The extension of conformal-map dynamics to multiply connected domains
is another mathematically challenging area, which has received some
attention recently but seems ripe for further development.
Understanding the exact solution structure of Laplacian-growth
problems using the mathematical abstraction of quadrature domain
theory holds great potential, especially given that mathematicians
have already begun to explore the extent to which the various
mathematical concepts extend to higher-dimensions (\cite{shap92}).
Describing multiply connected domains could pave the way for new
mathematical theories of evolving material microstructures. Topology
is the main difference between an isolated bubble and a dense
sintering compact. Microstructural evolution in elastic solids may be
an even more interesting, and challenging, direction for
conformal-mapping methods.

From a mathematical point of view, much remains to be done to place
stochastic conformal-mapping methods for interfacial dynamics on more
rigorous ground. This has recently been achieved in the simpler case
of Stochastic Loewner Evolution (SLE), which has a similar history to
the interfacial problems discussed here (\cite{kager04}). Oded Schramm
introduced SLE in 2000 as a stochastic version of the continuous
Loewner evolution from univalent function theory, which grows a
one-dimensional random filament from a disk or half plane. This
important development in pure mathematics came a few years after the
pioneering DLA papers of Hastings and Levitov in physics. A notable
difference is that SLE has a rigorous mathematical theory based on
stochastic calculus, which has enabled new proofs on the properties of
percolation clusters and self-avoiding random walks (in two
dimensions, of course). One hopes that someday DLA, DBM, ADLA, and
other fractal-growth models will also be placed on such a rigorous
footing.

Returning to materials applications, it seems there are many new
problems to be considered using conformal mapping. Relatively little
work has been done so far on void electromigration, viscous sintering,
solid pore evolution, brittle fracture, electrodeposition, and
solification in fluid flows. The reader is encouraged to explore these
and other problems using a powerful mathematical tool, which deserves
more attention in materials science.

\renewcommand\refname{\large Bibliography}

\bibliography{bazant3}

\begin{thebibliography}{61}
\expandafter\ifx\csname natexlab\endcsname\relax\def\natexlab#1{#1}\fi
\expandafter\ifx\csname url\endcsname\relax
  \def\url#1{{\tt #1}}\fi
\expandafter\ifx\csname urlprefix\endcsname\relax\def\urlprefix{URL }\fi

\bibitem[{Ball and Somfai(2002)}]{ball02}
Ball, R.~C. and Somfai, E., 2002.
\newblock Theory of diffusion controlled growth.
\newblock {\em Phys. Rev. Lett.\/} 89, 133503.

\bibitem[{Barra et~al.(2002{\natexlab{a}})Barra, Davidovitch and
  Procaccia}]{benny02}
Barra, F., Davidovitch, B. and Procaccia, I., 2002{\natexlab{a}}.
\newblock Iterated conformal dynamics and Laplacian growth.
\newblock {\em Phys. Rev. E\/} 65, 046144.

\bibitem[{Barra et~al.(2002{\natexlab{b}})Barra, Levermann and
  Procaccia}]{barra02}
Barra, F., Levermann, A. and Procaccia, I., 2002{\natexlab{b}}.
\newblock Quasistatic brittle fracture in inhomogeneous media and iterated
  conformal maps.
\newblock {\em Phys. Rev. E\/} 66, 066122.

\bibitem[{Batchelor(1967)}]{batchelor}
Batchelor, G.~K., 1967.
\newblock {\em An Introduction to Fluid Dynamics\/}.
\newblock Cambridge University Press.

\bibitem[{Bazant(2004)}]{bazant04}
Bazant, M.~Z., 2004.
\newblock Conformal mapping of some non-harmonic functions in transport theory.
\newblock {\em Proc. Roy. Soc. A\/} 460, 1433.

\bibitem[{Bazant et~al.(2003)Bazant, Choi and Davidovitch}]{ADLA}
Bazant, M.~Z., Choi, J. and Davidovitch, B., 2003.
\newblock Dynamics of conformal maps for a class of non-Laplacian growth
  phenomena.
\newblock {\em Phys. Rev. Lett.\/} 91, 045503.

\bibitem[{Ben~Amar(1999)}]{benamar99}
Ben~Amar, M., 1999.
\newblock Void electromigration as a moving free-boundary value problem.
\newblock {\em Physica D\/} 134, 275--286.

\bibitem[{Bensimon and Shraiman(1984)}]{ben84}
Bensimon, B. and Shraiman, D., 1984.
\newblock Singularities in non-local interface dynamics.
\newblock {\em Phys. Rev. A\/} 30, 2840--2842.

\bibitem[{Bunde and Havlin(1996)}]{bunde}
Bunde, A. and Havlin, S. (eds.), 1996.
\newblock {\em Fractals and Disordered Systems\/}.
\newblock 2nd edn. Springer, New York.

\bibitem[{Carrier et~al.(1966)Carrier, Krook and Pearson}]{carrier}
Carrier, G., Krook, M. and Pearson, C., 1966.
\newblock {\em Functions of a Complex Variable\/}.
\newblock McGraw-Hill, New York.

\bibitem[{Choi et~al.(2004{\natexlab{a}})Choi, Davidovitch and
  Bazant}]{choiADLA}
Choi, J., Davidovitch, B. and Bazant, M.~Z., 2004{\natexlab{a}}.
\newblock Crossover and scaling of Advection-Diffusion-Limited Aggregation.
\newblock In preparation.

\bibitem[{Choi et~al.(2004{\natexlab{b}})Choi, Margetis, Squires and
  Bazant}]{advdif}
Choi, J., Margetis, D., Squires, T.~M. and Bazant, M.~Z., 2004{\natexlab{b}}.
\newblock Steady advection-diffusion to finite absorbers in two-dimensional
  potential flows.
\newblock {\em J. Fluid Mech.\/} .

\bibitem[{Churchill and Brown(1990)}]{churchill}
Churchill, R.~V. and Brown, J.~W., 1990.
\newblock {\em Complex Variables and Applications\/}.
\newblock fifth edition edn. McGraw-Hill, New York.

\bibitem[{Crowdy(1999)}]{crowdy99}
Crowdy, D., 1999.
\newblock A note on viscous sintering and quadrature identities.
\newblock {\em Eur. J. Appl. Math.\/} 10, 623--634.

\bibitem[{Crowdy(2000)}]{crowdy00}
Crowdy, D., 2000.
\newblock Hele-Shaw flows and water waves.
\newblock {\em J. Fluid Mech.\/} 409, 223--242.

\bibitem[{Crowdy and Marshall(2004)}]{CroMar}
Crowdy, D. and Marshall, J., 2004.
\newblock Constructing multiply-connected quadrature domains.
\newblock {\em SIAM J. Appl. Math.\/} 64, 1334--1359.

\bibitem[{Crowdy and Tanveer(2004)}]{crowdy04}
Crowdy, D. and Tanveer, S., 2004.
\newblock The effect of finiteness in the Saffman-Taylor viscous fingering
  problem.
\newblock {\em J. Stat. Phys.\/} 114, 1501--1536.

\bibitem[{Crowdy(2003)}]{crowdy03}
Crowdy, D.~G., 2003.
\newblock Viscous sintering of unimodal and bimodal cylindrical packings with
  shrinking pores.
\newblock {\em Eur. J. Appl. Math.\/} 14, 421--445.

\bibitem[{Cummings et~al.(1999)Cummings, Hohlov, Howison and
  Kornev}]{cummings99}
Cummings, L.~M., Hohlov, Y.~E., Howison, S.~D. and Kornev, K., 1999.
\newblock Two-dimensional soldification and melting in potential flows.
\newblock {\em J. Fluid Mech.\/} 378, 1--18.

\bibitem[{Dai et~al.(1991)Dai, Kadanoff and Zhou}]{dai91}
Dai, W.-S., Kadanoff, L.~P. and Zhou, S.-M., 1991.
\newblock Interface dynamics and the motion of complex singularities.
\newblock {\em Phys. Rev. A\/} 43, 6672--6682.

\bibitem[{Davidovitch et~al.(1999)Davidovitch, Hentschel, Olami, Procaccia,
  Sander and Somfai}]{benny99}
Davidovitch, B., Hentschel, H. G.~E., Olami, Z., Procaccia, I., Sander, L.~M.
  and Somfai, E., 1999.
\newblock Diffusion-limited aggregation and iterated conformal maps.
\newblock {\em Phys. Rev. E\/} 59, 1368--1378.

\bibitem[{Duren(1983)}]{duren}
Duren, P.~L., 1983.
\newblock {\em Univalent Functions\/}.
\newblock Springer-Verlag, New York.

\bibitem[{Entov and Etingof(44)}]{entov91}
Entov, V.~M. and Etingof, P.~I., 44.
\newblock Bubble contraction in Hele-Shaw cells.
\newblock {\em Quart. J. Mech. Appl. Math\/} 507--535, 1991.

\bibitem[{Halsey(2000)}]{halsey00}
Halsey, T.~C., 2000.
\newblock Diffusion-limited aggregation: A model for pattern formation.
\newblock {\em Physics Today\/} 53, 36.

\bibitem[{Hastings(1997)}]{hastings97}
Hastings, M.~B., 1997.
\newblock Renormalization theory of stochastic growth.
\newblock {\em Phys. Rev. E\/} 55, 135.

\bibitem[{Hastings(2001)}]{hastings01}
Hastings, M.~B., 2001.
\newblock Fractal to nonfractal phase transition in the Dielectric Breakdown
  Model.
\newblock {\em Phys. Rev. Lett.\/} 87, 175502.

\bibitem[{Hastings and Levitov(1998)}]{hastings98}
Hastings, M.~B. and Levitov, L.~S., 1998.
\newblock Laplacian growth as one-dimensional turbulence.
\newblock {\em Physica D\/} 116, 244--252.

\bibitem[{Hermann and Roux(1990)}]{hermann}
Hermann, H.~J. and Roux, S. (eds.), 1990.
\newblock {\em Statistical Models for the Fracture of Disordered Media\/}.
\newblock North-Holland, Amsterdam.

\bibitem[{Hopper(1990)}]{Hopper}
Hopper, R., 1990.
\newblock Plane Stokes flow driven by capillarity on a free surface.
\newblock {\em J. Fluid Mech.\/} 213, 349--375.

\bibitem[{Howison(1986)}]{howison86}
Howison, S., 1986.
\newblock Fingering in Hele-Shaw cells.
\newblock {\em J. Fluid Mech.\/} 12, 439--453.

\bibitem[{Howison(1992)}]{howison92}
Howison, S.~D., 1992.
\newblock Complex variable methods in Hele-Shaw moving boundary problems.
\newblock {\em Euro. J. Appl. Math.\/} 3, 209--224.

\bibitem[{Jensen et~al.(2002)Jensen, Levermann, Mathiesen and
  Procaccia}]{jensen02}
Jensen, M.~H., Levermann, A., Mathiesen, J. and Procaccia, I., 2002.
\newblock Multifractal structure of the harmonic measure of diffusion-limited
  aggregates.
\newblock {\em Phys. Rev. E\/} 65, 046109.

\bibitem[{Kadanoff(1990)}]{kadanoff90}
Kadanoff, L.~P., 1990.
\newblock Exact solutions for the Saffman-Taylor problem with surface tension.
\newblock {\em Phys. Rev. Lett.\/} 65, 2986--2988.

\bibitem[{Kager and Nienhuis(2004)}]{kager04}
Kager, W. and Nienhuis, B., 2004.
\newblock A guide to Stochastic Loewner Evolution and its Applications.
\newblock {\em J. Stat. Phys.\/} 115, 1149--1229.

\bibitem[{Kornev and Mukhamadullina(1994)}]{kornev94}
Kornev, K. and Mukhamadullina, G., 1994.
\newblock Mathematical theory of freezing for flow in porous media.
\newblock {\em Proc. Roy. Soc. London A\/} 447, 281--297.

\bibitem[{Kruskal and Segur(1991)}]{kruskal91}
Kruskal, M. and Segur, H., 1991.
\newblock Asymptotics beyond all orders in a model of crystal growth.
\newblock {\em Stud. Appl. Math.\/} 85, 129.

\bibitem[{Levermann and Procaccia(2004)}]{levermann04}
Levermann, A. and Procaccia, I., 2004.
\newblock Algorithm for parallel Laplacian growth by iterated conformal maps.
\newblock {\em Phys. Rev. E\/} 69, 031401.

\bibitem[{Maclean and Saffman(1981)}]{maclean81}
Maclean, J.~W. and Saffman, P.~G., 1981.
\newblock The effect of surface tension on the shape of fingers in the
  Hele-Shaw cell.
\newblock {\em J. Fluid. Mech.\/} 102, 455.

\bibitem[{Muskhelishvili(1953)}]{musk}
Muskhelishvili, N., 1953.
\newblock {\em Some basic problems of the mathematical theory of elasticity\/}.
\newblock Noordhoff, Groningen, Holland.

\bibitem[{Needham(1997)}]{needham}
Needham, T., 1997.
\newblock {\em Visual Complex Analysis\/}.
\newblock Clarendon Press, Oxford.

\bibitem[{Niemeyer et~al.(1984)Niemeyer, Pietronero and Wiesmann}]{niemeyer84}
Niemeyer, L., Pietronero, L. and Wiesmann, H.~J., 1984.
\newblock Fractal dimension of dielectric breakdown.
\newblock {\em Phys. Rev. Lett.\/} 52, 1033--1036.

\bibitem[{Parisio et~al.(2001)Parisio, Moreas, Miranda and Widom}]{parisio01}
Parisio, F., Moreas, F., Miranda, J.~A. and Widom, M., 2001.
\newblock Saffman-Taylor problem on a sphere.
\newblock {\em Phys. Rev. E\/} 63, 036307.

\bibitem[{Richardson(1981)}]{Rich81}
Richardson, S., 1981.
\newblock Hele-Shaw flows with a free boundary produced by the injection of
  fluid into a narrow channel.
\newblock {\em J. Fluid Mech.\/} 56, 609--618.

\bibitem[{Richardson(1992)}]{Rich92}
Richardson, S., 1992.
\newblock Hele-Shaw flows with time-dependent free boundaries involving
  injection through slits.
\newblock {\em Stud. Appl. Math.\/} 87, 175--194.

\bibitem[{Richardson(2000)}]{Rich00}
Richardson, S., 2000.
\newblock Plane Stokes flow with time-dependent free boundaries in which the
  fluid occupies a doubly-connected region.
\newblock {\em Eur. J. Appl. Math.\/} 11, 249--269.

\bibitem[{Richardson(2001)}]{Rich01}
Richardson, S., 2001.
\newblock Hele-Shaw flows with time-dependent free boundaries involving a
  multiply-connected fluid region.
\newblock {\em Eur. J. Appl. Math.\/} 12, 571--599.

\bibitem[{Saffman(1959)}]{saffman59}
Saffman, P., 1959.
\newblock Exact solutions for the growth of fingers from a flat interface
  between two fluids in a porous medium.
\newblock {\em Q. J. Mech. Appl. Math.\/} 12, 146--150.

\bibitem[{Saffman and Taylor(1958)}]{saffman58}
Saffman, P.~G. and Taylor, G.~I., 1958.
\newblock The penetration of a fluid into a porous medium or Hele-Shaw cell
  containing a more viscous liquid.
\newblock {\em Proc. Roy. Soc. London A\/} 245, 312--329.

\bibitem[{Shapiro(1992)}]{shap92}
Shapiro, H., 1992.
\newblock {\em The Schwarz function and its generalization to higher
  dimension\/}.
\newblock Wiley, New York.

\bibitem[{Somfai et~al.(2003)Somfai, Ball and DeVita}]{somfai03}
Somfai, E., Ball, R.~C. and DeVita, J. P. ad~Sander, L.~M., 2003.
\newblock Diffusion-limited aggregation in channel geometry.
\newblock {\em Phys. Rev. E\/} 68, 020401.

\bibitem[{Somfai et~al.(1999)Somfai, Sander and Ball}]{somfai99}
Somfai, E., Sander, L.~M. and Ball, R.~C., 1999.
\newblock Scaling and crossovers in Diffusion Limited Aggregation.
\newblock {\em Phys. Rev. Lett.\/} 83, 5523.

\bibitem[{Stepanov and Levitov(2001)}]{stepanov01}
Stepanov, M.~G. and Levitov, L.~S., 2001.
\newblock Laplacian growth with separately controlled noise and anisotropy.
\newblock {\em Phys. Rev. E\/} 63, 061102.

\bibitem[{Tanveer(1993{\natexlab{a}})}]{tanveer93}
Tanveer, S., 1993{\natexlab{a}}.
\newblock Evolution of Hele-Shaw interface for small surface tension.
\newblock {\em Phil. Trans. Roy. Soc. London A\/} 343, 155--204.

\bibitem[{Tanveer(1993{\natexlab{b}})}]{tan93}
Tanveer, S., 1993{\natexlab{b}}.
\newblock Singularities in the classical Rayleigh-Taylor flow: formation and
  subsequent motion.
\newblock {\em Proc. Roy. Soc. A\/} 441, 501--525.

\bibitem[{Tanveer(2000)}]{surprise}
Tanveer, S., 2000.
\newblock Surprises in viscous fingering.
\newblock {\em J. Fluid Mech.\/} 409, 273--308.

\bibitem[{Varchenko and Etingof(1992)}]{var92}
Varchenko, A. and Etingof, P., 1992.
\newblock {\em Why the boundary of a round drop becomes a curve of order
  four\/}.
\newblock University Lecture Series, AMS, Providence.

\bibitem[{Wang and Suo(1997)}]{wang97}
Wang, W. and Suo, Z., 1997.
\newblock Shape change of a pore in a stressed solid via surface diffusion
  motivated by surface and elastic energy variations.
\newblock {\em J. Mech. Phys. Solids\/} 45, 709--729.

\bibitem[{Wang et~al.(1996)Wang, Suo and Hao}]{wang96}
Wang, W., Suo, Z. and Hao, T.-H., 1996.
\newblock A simulation of electromigration-induced transgranular slits.
\newblock {\em J. Appl. Phys.\/} 79, 2394--2403.

\bibitem[{Witten and Sander(1981)}]{witten81}
Witten, T.~A. and Sander, L.~M., 1981.
\newblock Diffusion-limited aggregation: a kinetic critical phenomenon.
\newblock {\em Phys. Rev. Lett.\/} 47, 1400--1403.

\bibitem[{Yoshikawa and Balk(1999)}]{yoshi99}
Yoshikawa, T. and Balk, A.~M., 1999.
\newblock The growth of fingers and bubbles in the strongly nonlinear regime of
  the Richtmyer-Meshkov instability.
\newblock {\em Phys. Lett. A\/} 251, 184--190.

\bibitem[{Zakharov(1968)}]{zakharov68}
Zakharov, V.~E., 1968.
\newblock Stability of periodic waves of finite amplitude on the surface of
  deep fluid.
\newblock {\em J. Appl. Mech. Tech. Phys.\/} 2, 190.

\end{thebibliography}

\begin{flushright}
Martin Z. Bazant\\
Department of Mathematics\\
Massachusetts Institute of Technology\\
Cambridge, MA USA\\
bazant@mit.edu\\
\ \\
Darren Crowdy\\
Department of Mathematics\\
Imperial College\\
London, United Kingdom\\
d.crowdy@imperial.ac.uk
\end{flushright}

\end{document}